\documentclass[fleqn,usenatbib]{mnras}

\usepackage{newtxtext,newtxmath}

\usepackage[T1]{fontenc}

\DeclareRobustCommand{\VAN}[3]{#2}
\let\VANthebibliography\thebibliography
\def\thebibliography{\DeclareRobustCommand{\VAN}[3]{##3}\VANthebibliography}

\usepackage{graphicx}	
\usepackage{amsmath}	
\usepackage{longtable}

\title[FU Ori outburst WTP\,10aaauow]{WTP\,10aaauow: Discovery of a new FU Ori outburst towards the RCW\,49 star-forming region in NEOWISE data}

\author[V. Tran et al.]{
Vinh Tran,$^{1}$\thanks{E-mail: vinhtran@mit.edu}
Kishalay De,$^{1,2}$
Lynne Hillenbrand$^{3}$
\\
$^{1}$MIT-Kavli Institute for Astrophysics and Space Research, 77 Massachusetts Ave., Cambridge, MA 02139, USA\\
$^{2}$NASA Einstein Fellow\\
$^{3}$Department of Astronomy, California Institute of Technology, Pasadena CA 91125\\
}

\date{Accepted XXX. Received YYY; in original form ZZZ}

\pubyear{2015}

\begin{document}
\label{firstpage}
\pagerange{\pageref{firstpage}--\pageref{lastpage}}
\maketitle

\begin{abstract}
Large-amplitude accretion outbursts in young stars are expected to play a central role in proto-stellar assembly. Outburst identification historically has taken place using optical techniques, but recent, systematic infrared searches are enabling their discovery in heavily dust-obscured regions of the Galactic plane. Here, we present the discovery of WTP\,10aaauow, a large-amplitude mid-infrared (MIR) outburst identified in a systematic search of NEOWISE data using new image subtraction techniques. The source is located towards the RCW\,49 star-forming region, and estimated to be at a distance of $\approx 4$\,kpc via Gaia parallax measurement. Concurrent with the MIR brightening, the source underwent a $\gtrsim5$\,mag outburst in the optical and near-infrared (NIR) bands, reaching a peak luminosity of $\approx260$\,L$_\odot$ in 2014-2015, followed by a slow decline over the next 7 years. Analysis of the pre- and post-outburst spectral energy distributions reveal a pre-outburst stellar photosphere at a temperature of $3600-4000$\,K, surrounded by a likely two-component dust structure similar to a flat-spectrum or Class I type YSO. We present optical and NIR spectroscopy that show a GK-type spectrum in the optical bands exhibiting complex line profiles in strong absorption features, and evidence for a wind reaching a terminal velocity of $\approx 400$\,km\,s$^{-1}$. The NIR bands are characterized by a cooler M-type spectrum exhibiting a forest of atomic and molecular features. All together, the spectra demonstrate that WTP\,10aaauow is an FU Ori type outburst. Ongoing systematic infrared searches will continue to reveal the extent of this population in the Galactic disk. 
\end{abstract}

\begin{keywords}
methods: observational -- stars: formation -- ISM: H\,II regions
\end{keywords}



\section{Introduction}

Star formation proceeds through the gravitational collapse of molecular cloud cores to form a central protostar, initially through direct spherical accretion, followed by disk accretion \citep{Armitage2019}. The process of accretion transports both mass and angular momentum throughout the proto-planetary disk \citep{pringle1981}, and is an intrinsically unsteady process \citep{turner2014}.  Observationally, variability is omnipresent in young stellar objects (YSOs); see \cite{fischer2023} for a review.  Minor effects such as time-varying dust extinction, irregular magnetospheric accretion, or surface inhomogeneites on rotating stellar photospheres can cause low-amplitude variability on short timescales. Observed variability at higher-amplitudes can also be extinction-driven or accretion-driven.  Of particular relevance here are episodic brightening events that are both large-amplitude and long-lasting.  Such outbursts are related to substantial changes in the accretion rate from the disk to the central protostar. 

Outbursts appear to play a crucial role in stellar mass assembly \citep[][and references therein]{vorobyov2021} as well as affecting the details of the earliest stages of stellar evolution \citep{hosokawa2011}. The largest amplitude outbursts, known as FU Ori objects, exhibit luminosity increases $\gtrsim 100\times$ lasting for decades to centuries, and are a well-recognized mechanism for stellar growth \citep{hartmann1996}. 
Lower-amplitude outbursts such as the EX Lup-type \citep{Herbig1977} are more frequent but contribute less substantially to mass gain. Commonly discussed outburst mechanisms include: larger-scale phenomena such as binary disruption of the disk and instabilities due to infall from the envelope, to intermediate-scale effects such as planet-induced instabilities, various flavors of gravo-magneto instabilities, or thermal instability related to disk opacity, and finally to smaller-scale instabilities within the star-disk interaction region or the magnetosphere \citep[see references in e.g.][]{armitage2015,auduard2014,vorobyov2021,nayakshin2024b}.  It is likely that different physical scenarios operate or may be dominant in different FU Ori-type accretion/outflow systems.  Also possible are cascades of these effects \citep{skliarevskii2024}.  

The small population of known FU Ori sources suggests that they may be extremely rare. Only around 14 FU Ori outbursts have been recorded as they have occurred, and are also validated with both optical and infrared spectra indicating a luminous accretion disk. Another 13 sources have been identified as outbursters post facto, via their convincing spectral signatures \citep[e.g.][]{hartmann1996, Connelley2018, Contreras-Pena2019, nagy2023, hillenbrand2023}. New candidates that have incomplete evidence continue to be uncovered \citep{Guo2021, Park2021, Zakri2022}. The diversity of the FU Ori class was illustrated in the lightcurves and some spectral properties of just its first three members \citep{clarke2005}. An increased sample of candidates has increased that diversity, though some families of behavior are beginning to emerge.

Observationally, FU Ori outbursts are characterized by brightness increases $\gtrsim 3.5-5$\,magnitudes and the spectral characteristic of GK-type giants/supergiants in the optical bands \citep{Herbig1977}, transitioning to cooler M-type giants/supergiants in the near-infrared (NIR) bands \citep{Connelley2018}. The wavelength-dependent spectral types can be explained with accretion disk models including a temperature and velocity gradient with radius \citep{Kenyon1988, Welty1992}. Although optical time domain surveys have substantially advanced the field towards systematic discovery and characterization, our empirical understanding of the distribution of the amplitudes, durations, and duty cycles of the largest outbursts remains highly uncertain \citep[see][for review]{fischer2023}. 

Infrared searches for FU Ori outbursts provide a unique opportunity to discover and develop a census of these events due to their sensitivity towards highly dust-obscured star-forming regions. Ground-based searches in the NIR have already begun to reveal a new window into YSO variability -- both individual large-amplitude outbursts (e.g. \citealt{Hillenbrand2021}) amenable for detailed infrared characterization, as well as populations of smaller amplitude variables associated with young stellar variability (e.g. \citealt{Contreras-Pena2017, Guo2022}). With its all-sky coverage, mid-infrared (MIR) sensitivity and long/uniformly sampled temporal baseline, the Wide-field Infrared Survey Explorer (WISE) provides a unique dataset to search for large-amplitude, long-duration FU Ori outbursts, towards developing a complete census of these events near the dust obscured Galactic plane where shorter wavelength searches become ineffective.

To this end, our team has begun a systematic search for large-amplitude MIR outbursts utilizing image subtraction techniques on WISE images. In this paper, we present a new addition to the FU Ori class discovered in our search, named WTP\,10aaauow. Section \ref{sec:discovery} describes our discovery of the source and subsequent follow-up observations. We accumulate archival information on the source in Section \ref{sec:archive}, and analyze its properties in Section \ref{sec:analysis}. We conclude with a discussion of the source properties in Section \ref{sec:discussion}.

\section{Discovery and Observations}
\label{sec:discovery}
\subsection{Outburst Discovery in NEOWISE Data}
The Wide-field Infrared Survey Explorer (WISE) satellite \citep{Wright2010}, re-initiated as the NEOWISE mission \citep{Mainzer2014}, has been carrying out an all-sky MIR survey in the $W1$ ($3.4$\,$\mu$m) and $W2$ ($4.6$\,$\mu$m) bands since 2014. In its ongoing survey, NEOWISE revisits each part of the sky once every $\approx 0.5$\,yr. 
We have carried out a systematic search for transients in time-resolved coadded images created as part of the unWISE project \citep{Lang2014,Meisner2018}, the details of which will be presented in De et al. (in prep). In brief, we used a customized code \citep{De2019} based on the ZOGY algorithm \citep{Zackay2016} to perform image subtraction on the NEOWISE images using the co-added images of the WISE mission (obtained in 2010-2011) as reference images\footnote{For all transients identified in the WISE Transient Pipeline (WTP, De in prep.), we adopt the naming scheme WTP\,XXYYYYYY, where XX indicates the year of first detection and YYYYYY is a six letter alphabetical code.}. Our pipeline produces a database of all transients down to a statistical significance of $\approx 10\sigma$. Follow-up for the sources was coordinated using the \texttt{fritz} astronomical data platform \citep{vanderWalt2019}. 

\begin{figure*}
    \centering
    \includegraphics[width=\textwidth]{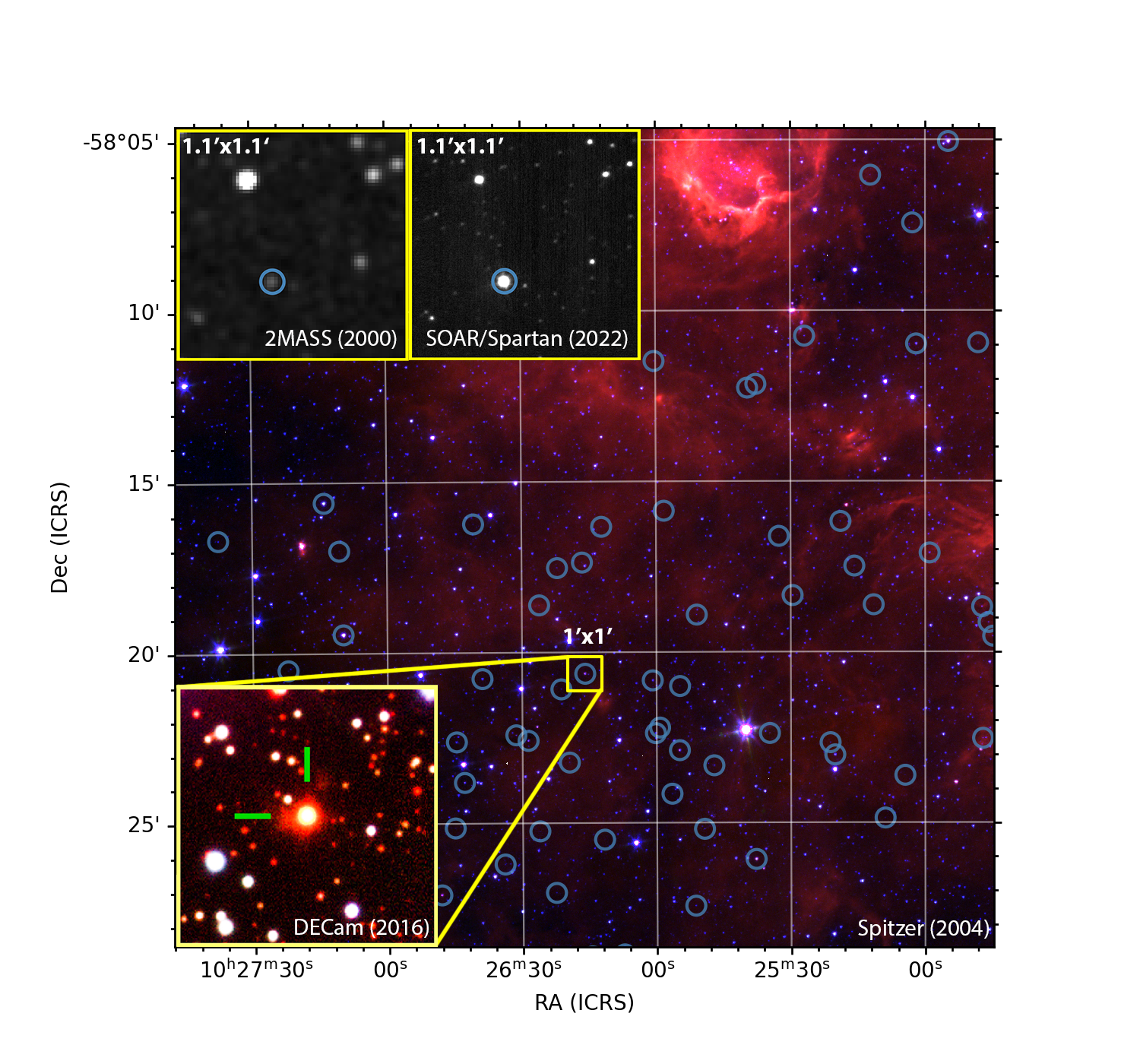}
    \caption{Color composite of the $24\arcmin \times 24\arcmin$ region surrounding WTP\,10aaauow constructed from Spitzer Space Telescope \citep{Werner2004} images taken with IRAC Channel 4 (8 $\mu$m) as red, Channel 3 (5.8 $\mu$m) as green, and Channel 2 (4.5 $\mu$m) as blue. The nebulosity in the northwest part belongs to the H\,II region RCW\,49, and the blue circles mark potential YSOs in the area (see Section \ref{sec:YSOs}).  The zoom-in yellow inset on the bottom left shows an optical image of the $1\arcmin \times 1\arcmin$ region surrounding WTP\,10aaauow, created using $z$ band (926 nm; red), $r$ band (642 nm; green), and $g$ band (473 nm; blue) images taken from the DECam Galactic Plane Survey \citep{Schlafly2018}. There is nebulosity clearly visible surrounding the bright transient (highlighted with the green cross-hair). The two insets on the top left depict the pre-outburst (left) and post-outburst (right) $K$ band images of the region from 2MASS (in 2000) and the Spartan camera (in 2022), depicting the NIR brightening of the transient (circled in blue).}
    \label{fig:color}
\end{figure*}

\begin{figure*}
    \centering
        \includegraphics[width=\textwidth]{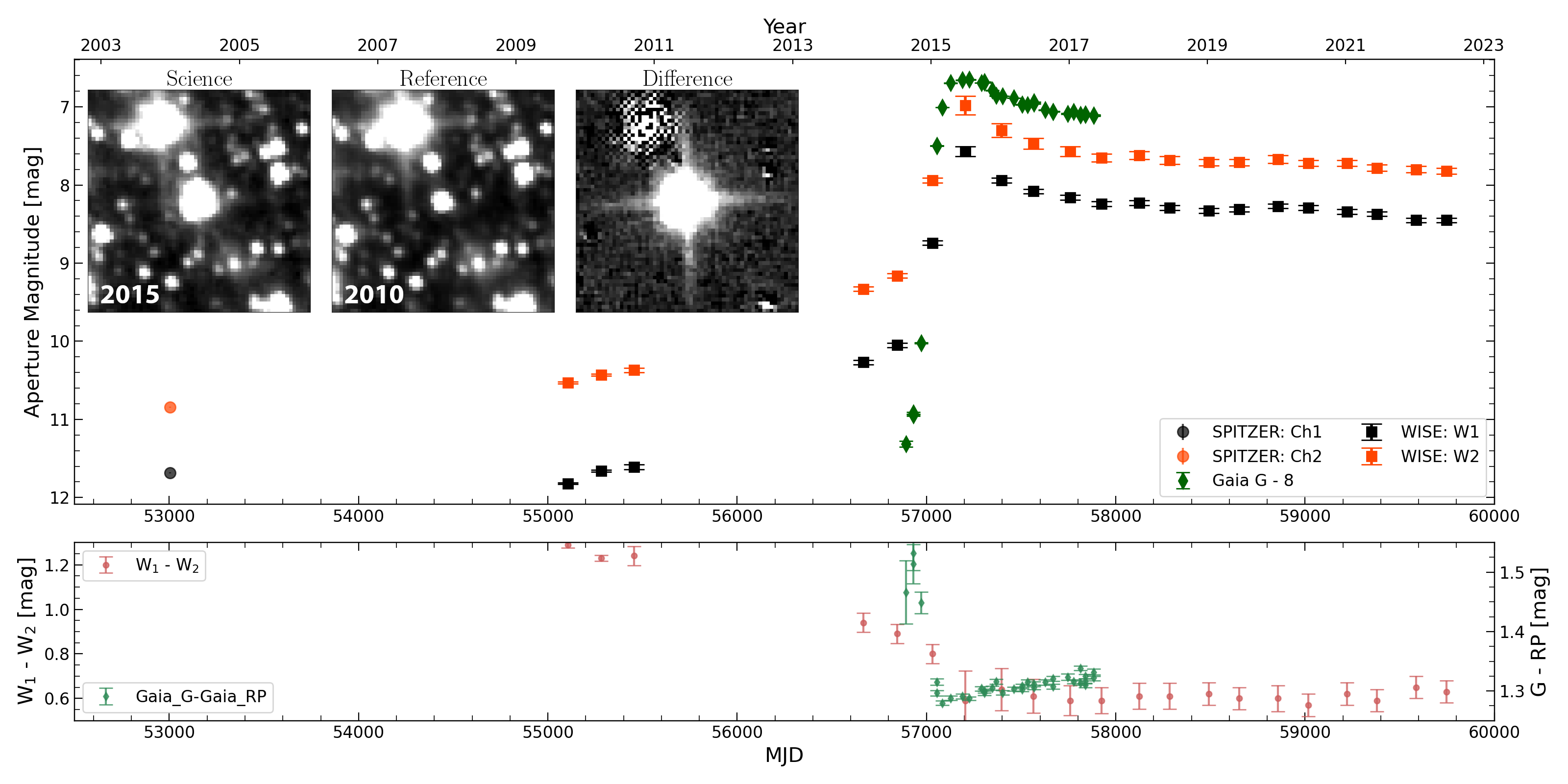}
    \caption{{\it Top:} MIR light curve of WTP\,10aaauow obtained from WISE photometry (supplemented by Spitzer data), together with optical photometry from Gaia DR3 (shifted by -8\,mag, as indicated in the legend). The upper left inset shows a triplet of the science (taken in 2015), reference (taken in 2010), and difference image of the location obtained from our difference imaging pipeline, clearly showing the MIR brightening of WTP\,10aaauow. {\it Bottom:} The MIR $W1-W2$ and the visual G-RP color evolution of WTP\,10aaauow.}
    \label{fig:lc}
\end{figure*}

We identified the transient source WTP\,10aaauow 
as a large-amplitude outburst at J2000 coordinates $\alpha=$10:26:15.98, $\delta=-$58:20:37.80, near the H II region RCW~49 and the Carina nebula (Figure \ref{fig:color}). The transient is coincident with a source in archival AllWISE images, exhibiting a shallow brightening phase of $\approx 1$\,mag between 2010 and 2014 (Figure \ref{fig:lc}), followed by a sudden brightening of $\gtrsim 3$\,mags between $2014$ and $2015$. The corresponding brightening rates are $\approx 0.02$\,mag\,month$^{-1}$ in the early shallow brightening phase and $\approx 0.25$\,mag\,month$^{-1}$ in the subsequent rapid brightening phase. The transient has since exhibited a slow fading, declining in flux at a rate of $\approx 0.01$\,mag\,month$^{-1}$ between 2015 and 2022.

Although the outburst was first detected in the pipeline during the early brightening phase, the transient is formally saturated in the unWISE images near the peak of the outburst. Therefore, we derive the MIR light curve of the source by performing the recommended bias corrections for saturated sources in the NEOWISE single epoch catalogs\footnote{We note that although there is a faint source $\approx 7$\arcsec away from the outburst detected in archival Spitzer images, it does not contaminate the source photometry (as confirmed by the consistency of the WISE and Spitzer photometry prior to outburst; Figure \ref{fig:lc}) as the WISE Point Spread Function is $<5$\arcsec and the nearby source is $\lesssim 20$\% of the pre-outburst progenitor flux.} as explained in the WISE Data Release document\footnote{\url{https://wise2.ipac.caltech.edu/docs/release/neowise/expsup/sec2_1civa.html}}. We use standard aperture photometry for epochs prior to the start of NEOWISE (when the source was faint). 
The complete WISE light curve of the source after correction of the instrumental effects and conversion to the AB magnitude system is shown in Figure \ref{fig:lc}. 


\subsection{Follow-up Imaging}

\begin{table}
    \centering
    \begin{tabular}{cccc}
        \hline
        \hline
        Instrument & MJD & Magnitude & Band \\
        \hline
        LCO/Sinistro & 59965.32 & $19.35\pm0.01$ & g  \\
        LCO/Sinistro & 59965.32 & $16.65\pm0.01$ & r \\
        LCO/Sinistro & 59965.33 & $15.19\pm0.02$ & i \\
        \hline
        Gaia & 56892.76 & $19.32\pm0.04$ & G \\
        Gaia & 56892.76 & $21.48\pm0.63$ & BP \\
        Gaia & 56892.76 & $17.85\pm0.04$ & RP \\
        Gaia & 56929.55 & $18.94\pm0.01$ & G \\
        Gaia & 56929.55 & $17.41\pm0.03$ & RP \\
        Gaia & 56929.55 & $24.57\pm7.65$ & BP \\
        ... & ... & ... & ... \\
        \hline
    \end{tabular}
        \caption{ Photometric measurements of WTP\,10aaauow. Only a representative sample of the first lines are presented. The full table is shown in Appendix \ref{sec:appendix}.}
    \label{tab:photo_truncated}
\end{table}

We obtained one epoch of multi-color NIR imaging of WTP\,10aaauow on UT 2022-08-04, using the Spartan camera \citep{Loh2012} on the 4.1 m Southern Astrophysical Research (SOAR) Telescope as part of program SOAR 2022B-005 (PI: De). We obtained a series of 10 dithered exposures of the field with exposure times of 30\,s, 20\,s, and 10\,s in the $J$, $H$, and $Ks$ filters respectively. The data were detrended followed by astrometric and photometric calibration against nearby 2MASS sources using a modified version of the imaging pipeline described in \citet{De2019}. Performing aperture photometry at the position of the transient, we measure NIR fluxes (all in the Vega magnitude system) of $J = 11.50 \pm 0.01$\,mag, $H = 10.20 \pm 0.01$\,mag and $Ks = 9.24 \pm 0.01$\,mag. 

We obtained follow-up optical imaging of the transient using the Sinistro imager on the 1\,m Las Cumbres Observatory network (\citealt{Brown2013}; Program  CON2022B-011). The data were acquired on UT 2023-01-21 in the $gri$ filters for a total exposure time of 900\,s each. The data were astrometrically calibrated and stacked using a custom reduction pipeline \citep{De2020}, and photometric calibration was performed against photometric calibrators in the field from the Legacy Imaging Survey \citep{Dey2019} in the AB magnitude system. We measure $g = 19.35 \pm 0.01$\,mag, $r = 16.65\pm 0.01$\,mag and $i = 15.19 \pm 0.02$\,mag for the transient at this epoch. 

\subsection{Follow-up Spectroscopy}

We obtained follow-up NIR spectroscopy of WTP\,10aaauow
to confirm the classification of the transient. We obtained one epoch of NIR ($\approx 1.0 - 2.5\,\mu$m) spectroscopy using the Folded Port Infrared Echellette (FIRE; \citealt{Simcoe2013}) on the 6.5\,m Magellan Baade Telescope on UT 2022-05-25 and 2022-06-15. Observations in the first epoch were obtained in the low resolution prism mode ($R \approx 100$) using the 0.6\,\arcsec slit, over a set of dithered exposures amounting to a total exposure time of 300\,s. The second epoch of spectroscopy was carried out in the echelle mode ($R \approx 5000$) to obtain higher resolution confirmation of the features detected in the first epoch. The echelle mode observations were carried out using the same 0.6\,\arcsec slit over a set of dithered exposures amounting a total exposure time of $\approx 1200$\,s. For each observation, we also observed a nearby telluric standard star for calibration. The data were reduced and extracted using the \texttt{firehose} pipeline\footnote{\url{https://wikis.mit.edu/confluence/display/FIRE/FIRE+Data+Reduction}} for both the prism and echelle mode data, followed by flux calibration and telluric correction using the \texttt{xtellcor} package \citep{Vacca2003}. 

We obtained a red optical spectrum of WTP\,10aaauow on UT 2023-01-11, using the Goodman High Throughput Spectrograph \citep{Clemens2004} on SOAR as part of program SOAR 2022B-005 (PI: De). The data were acquired with the 600 lines/mm grating (630 - 893 nm; $R \simeq 1400$), for a total exposure time of 1200\,s. Data reduction and flux calibration with a spectro-photometric standard were performed using the \texttt{pypeit} package \citep{Prochaska2020}.

\subsection{Archival Information}
\label{sec:archive}

\subsubsection{Survey Imaging}
The sky region near WTP\,10aaauow has been previously observed in both the optical and NIR bands. In the optical, no source is detected in the pre-outburst Digitized Sky Survey (DSS) images from the 1980s. Querying the DSS catalog within a region of $10$\arcmin, we estimate the corresponding limiting magnitude to be $\gtrsim 17.5$\,mag. Additionally, no source has been detected by the All-Sky Automated Survey for Supernovae \citep{Shappee2014,Kochanek2017} pre- and post-outburst with the detection limit of $\sim$16\,mag and $\sim$17\,mag in the V and g filters, respectively. This is consistent with the $\sim$16.5\,mag post-outburst photometry data of Gaia BP band, which has a similar passband to that of traditional V band. However, there is a source cataloged in the IPHaS/VPHaS$+$ \citep{vphas} Galactic plane survey, with faint-state photometry reported from 2014 of $r = 21.99\pm0.21$; $i = 19.89\pm 0.07$. There is also an $H\alpha$ measurement of $20.70 \pm 0.17$, which suggests the source was a strong emission-line object in its quiescent state. In post-outburst imaging, a red source with extended nebulosity (Figure \ref{fig:color}) is clearly detected in the DECam Galactic plane survey \citep{Schlafly2018}. Compared to archival DECam photometry taken at MJD $\approx 57590$ (close to the peak of the outburst in Figure \ref{fig:lc}), the source had faded by $\approx 1.2$\,mag by the epoch of our optical imaging. 

In the infrared, a point source is clearly detected during the quiescent phase in archival images from 2MASS\footnote{ \url{https://irsa.ipac.caltech.edu/data/2MASS/docs/releases/allsky/doc/sec2_2b.html}} \citep{Skrutskie2006}, near the completeness limits of the survey\footnote{The 2MASS catalogs are complete to $J \approx 16.4$\,mag, $H\approx15.5$\,mag and $Ks \approx 14.8$\,mag over $50$\% of the sky.}, and from AllWISE \citep{Wright2010}. Comparing to archival 2MASS magnitudes, the recent NIR magnitude measurements (which are likely $\approx 1-2$\,mags fainter than peak outburst flux  based on comparison of the current MIR magnitude compared to MIR peak; Figure \ref{fig:lc}) suggest the source has exhibited $\gtrsim 5$\,mag brightening in the NIR. The source was also observed by Spitzer/IRAC in four channels as part of the GLIMPSE survey \citep{Benjamin2003}, and we obtained the photometry from the online Spitzer source catalog \citep{Spitzer2021}. The MIR measurements are shown in Figure~\ref{fig:lc}.

\subsubsection{Catalog Data}

WTP\,10aaauow has appeared in a number of YSO catalogs, namely \cite{vioque2020} and \cite{kuhn2021}. In terms of source properties, Gaia DR3 \citep{GaiaCollaboration2022} reports a parallax $\pi = 0.25\pm0.03$\,mas (corresponding to distance $d= 4.0 \pm 0.4$ kpc); proper motions $\mu_{RA} = -5.08\pm0.04$\,mas\,yr$^{-1}$ and $\mu_{Dec.}= 3.10\pm0.03$\,mas\,yr$^{-1}$; and radial velocity of $-62.57\pm 5.14$\,km\,s$^{-1}$. Elsewhere in literature catalogs, \cite{fouesneau2022} derived extinction values, finding $A_0 = 4.6$ mag as the ``best" value and a $A_0 = 3.6$ mag as the median value in their posterior. Due to the nature of their methods and the variable nature of the source, these estimates could be unreliable, but $\approx 5$\,mag of visual extinction appears reasonable for the source. However, we do not consider any of the stellar parameters derived for WTP\,10aaauow as viable.

Although no objects at the position of WTP\,10aaauow appear in the Gaia Alerts service \citep{gaia_alerts}, the source does have a well-sampled optical wavelength light curve in Gaia DR3 \citep{GaiaCollaboration2022}.
As illustrated in Figure~\ref{fig:lc}, Gaia captured the source brightening beginning in late 2014 to its peak in early 2015, and the subsequent plateau phase. Coincident with the brightening, the source became bluer in its optical color by several tenths of a magnitude, as it does in its MIR color. All photometric measurements of WTP 10aaauow, including Gaia measurements, are shown in Table \ref{tab:photo_truncated}.

\section{Analysis and Results}
\label{sec:analysis}
In this section, we present evidence that WTP\,10aaauow is an outbursting young stellar object of the FU Orionis type.
This conclusion is based on i) its proximity to a large star-forming region with hundreds of YSOs within a few tens of arcmin, ii) its spectral energy distribution indicating a large dust envelope, iii) large-amplitude outburst and high peak luminosity, and iv) its spectroscopic signatures which are consistent with both stellar youth and a characteristic wavelength-dependent absorption spectrum for the FU Ori class.

\subsection{Identification of Nearby Candidate Young Stars}
\label{sec:YSOs}

WTP\,10aaauow is located around 35\arcmin\ southeast of the H\,II region RCW\,49 (or NGC\,3247), home to several hundred young stellar object candidates as well as a hub of occurring star-formation \citep{Whitney2004}. This region is situated around 4.2\,kpc from the Solar system \citep{Churchwell2004}, which is very similar to the distance of WTP\,10aaauow directly reported in Gaia ($4.0\pm0.4$\,kpc). Using this distance, we calculate the physical separation between our object and the region to be $\approx 40$\,pc, placing WTP\,10aaauow near the edge of the H II region. There are no nearby candidate YSOs within 2\arcmin\ of WTP\,10aaauow according to SIMBAD, so we apply a color cut to the Spitzer/GLIMPSE \citep{Churchwell2009} and 2MASS \citep{Cutri2003, Skrutskie2006} source catalogs in the sky region to identify nearby candidate YSOs. We follow the process detailed in \cite{Gutermuth2009} to eliminate IR-excess contaminants, such as star-forming galaxies, broad-line active galactic nuclei, and shock emission regions. Afterward, we use a simpler color cut of $[3.6]-[4.5] > 0.15$ and $J-K > 0.8$ to find IR-excess sources that are likely YSOs, resulting in 176 potential YSOs within 20\arcmin\ from the source. 




The candidate YSO sources, including our source, are shown in Figure \ref{fig:color}, with blue circles marking their positions. The closest of the potential YSOs to our source is SSTSL2 J102621.26-582104.8 followed by SSTSL2 J102600.85-582049.8 with separations of 50\arcsec\ and 119\arcsec\, respectively. While these are the only YSO candidates within 2\arcmin\ ($\approx 2.3$\,pc at the distance of WTP\,10aaauow) of the source, there are an additional 12 sources within 4\arcmin. Out of the 176 potential YSOs, there are 16 sources with distances measured by Gaia DR3 \citep{GaiaCollaboration2022}. 11 of these sources are situated between 3.2 and 4.7\,kpc, making these sources physically close to WTP\,10aaauow. 110 out of the 176 sources have proper motions detected in Gaia, and their distribution is well described by a Gaussian function with (mean, standard deviation) of (-5.70, 1.38) and (3.20, 0.98) mas\,yr$^{-1}$ in the RA and Dec directions respectively, very similar to that reported for WTP\,10aaauow ($\mu_{RA} = -5.08\pm0.04$\,mas\,yr$^{-1}$; $\mu_{Dec.}= 3.10\pm0.03$\,mas\,yr$^{-1}$). Based on the analysis above, we speculate that the majority of the 176 potential YSOs in the region are possibly proximal to the source. We interpret the existence of the large number of nearby YSO candidates as evidence in favor of the classification of WTP\,10aaauow as a YSO in the same star-forming region.




\subsection{Spectral Energy Distribution (SED)}

\begin{figure}
    \centering
        \includegraphics[width=0.48 \textwidth]{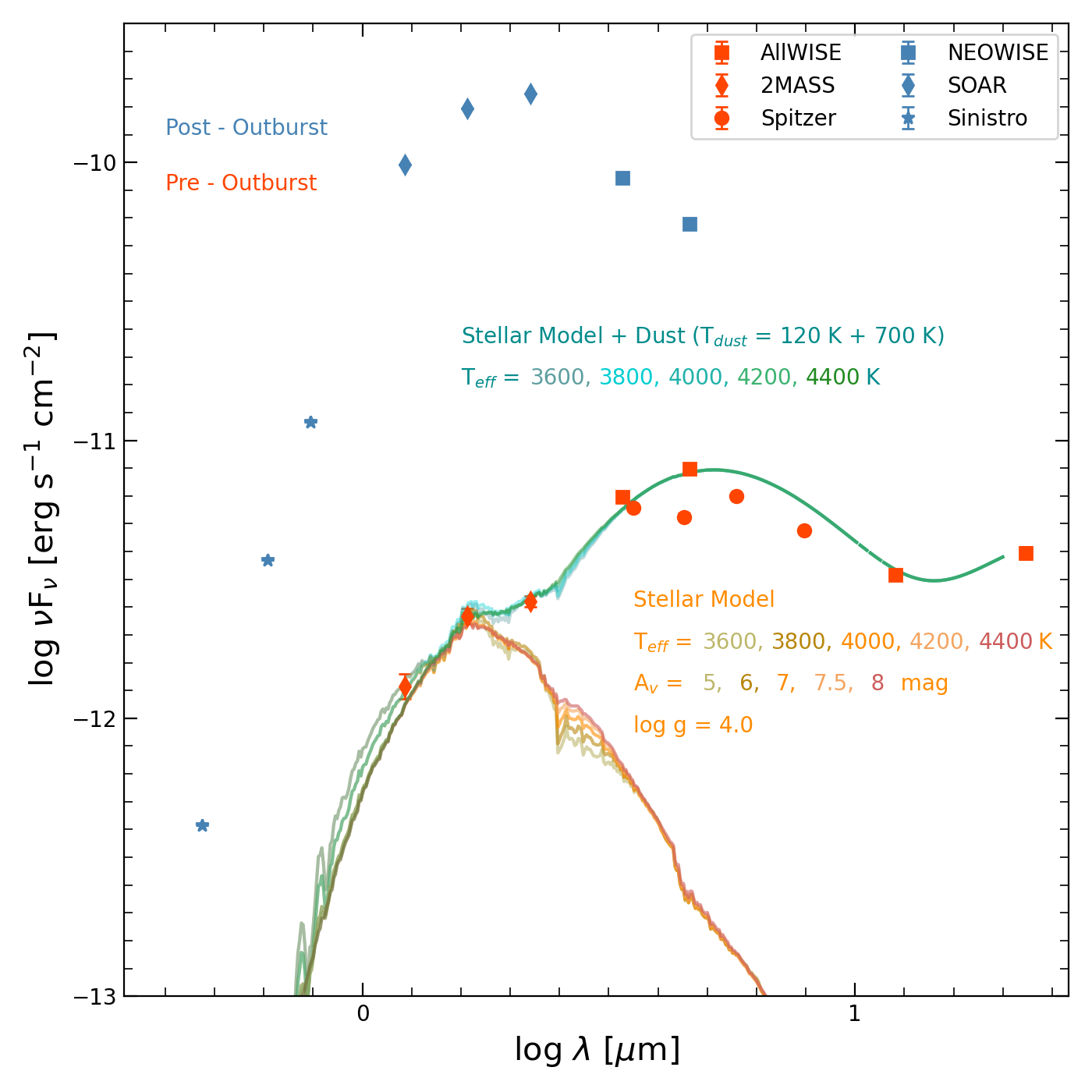}
    \caption{The post- (blue) and pre-outburst (orange) spectral energy distribution of WTP\,10aaauow. Reddened photosphere models with effective temperatures of $3600-4400$\,K and visual extinction of $5-8$ magnitudes are shown in orange-tinted lines. 
    The pre-outburst photosphere models are normalized to the pre-outburst $J$, $H$ bands. In addition, two black-body spectral energy distributions with effective temperatures of 120\,K and 700\,K are added to the photosphere models and shown as blue-tinted lines to explain the IR excess at longer wavelengths.}
    \label{fig:sed}
\end{figure}

Figure \ref{fig:sed} shows the $0.5-22$\,microns photometric measurements of WTP\,10aaauow pre- and post-outburst. The pre-outburst data comprises of AllWISE $W1$, $W2$, $W3$, $W4$ bands ($\lambda$ = 3.4, 4.6, 12, 22\,$\mu$m); Spitzer/IRAC channel 1, 2, 3, 4 ($\lambda$ = 3.6, 4.5, 5.8, 8\,$\mu$m); and 2MASS $J$, $H$, $K$ bands ($\lambda$ = 1.2, 1.6, 2.2\,$\mu$m). However, it must be noted that these measurements are taken at different times, in 2010, 2004, and 2000, respectively, so the compiled SED is not derived from strictly contemporaneous measurements.  Additionally, Gaia data are not used in the SED since, as shown in Figure \ref{fig:lc}, 
the earliest measurement from Gaia likely captured WTP\,10aaauow while it was already undergoing the outburst. In contrast, the post-outburst data from NEOWISE $W1$, $W2$ band ($\lambda$ = 3.4, 4.6\,$\mu$m); SOAR/Spartan $J$, $H$, $K$ band ($\lambda$ = 1.24, 1.63, 2.15\,$\mu$m); and Sinistro $g$, $r$, $i$ band ($\lambda$ = 477, 622, 755\,nm) are all measured within half a year, indicating that the assembled post-outburst SED is a viable estimate at 7 years after the outburst.

We use photospheric models from NextGen2 \citep{Hauschildt1999} over a range of visual extinctions to fit the $J+H$ band region of the pre-outburst SED. Additionally, we use two black-body SEDs to fit the excess in the redder NIR and MIR region as a simple illustration of the temperatures involved. The excess is most likely due to the presence of an extended dust envelope surrounding an accretion disk that we do not model in detail. We choose the temperatures for the black-body SEDs based on the peak near WISE $W1$ and $W2$ band, as well as the excess near WISE $W4$ band, resulting in T$_{dust} \approx$ 700\,K and 120\,K (noting that the temperature of the colder component is poorly constrained due to the limited wavelength coverage). We find that the pre-outburst data can be explained by photospheric models as cold as 1000\,K and hot as 8000\,K over a wide range of extinction, dust temperature, and black-body SEDs. 

While the total integrated Galactic extinction along this line-of-sight has been measured to be $A_V \approx 7.3$\,mag \citep{Schlafly2011}, a more detailed estimate of the extinction to WTP\,10aaauow can be made by comparing its near-infrared SED to that of other FU Ori objects.  Adopting the method of \cite{ashraf2023} who compared extinction and mid-infrared W1-W2 colors, we find $A_V \approx 4-6$\,mag. Following the procedure of \cite{Connelley2018} to deredden the near-infrared spectrum of Gaia\,17bpi which has the extinction of $3-3.8$\,mag \citep{Hillenbrand2018,Carvalho2022,Rodriguez2022}, we find $A_V = 6-7$\,mag. Extinction is also informed by the $A_V \approx$ 5 mag found by \cite{fouesneau2022} from optical data. Although somewhat uncertain, these independent extinction estimates constrain the effective temperature of the stellar model to be in the range $3600 - 4000$\,K. 

The corresponding SED models are shown in Figure \ref{fig:sed}. Empirically, the spectral index over the standard 2 to 24 $\mu$m range \citep{greene1994} of the progenitor source is $\alpha = +0.14$, classifying WTP\,10aaauow as a Class I or flat spectrum type YSO. While the SPICY catalog \citep{kuhn2021} reports a spectral index of $\alpha = -0.31$, the difference arises due to their differing definition of the spectral index based on the $4.5\,\mu$m to $8.0\,\mu$m or $W4$ color; however Figure \ref{fig:sed} clearly demonstrates the rising spectrum of the progenitor to the mid-infrared bands confirming a class I or flat spectrum classification.

\subsection{Outburst Luminosity}

Integrating the SED from Figure \ref{fig:sed}, we find the post-outburst luminosity of WTP\,10aaauow, measured at $\approx 7$\,years post-outburst to be $\approx 110$\,L$_\odot$ for a distance of 4\,kpc. However, the luminosity at outburst peak is expected to have been higher. From Figure \ref{fig:lc}, we notice that in the  $W1$ and $W2$ bands, the current magnitudes dim by $\approx 0.9$ mag compared to the corresponding peak magnitudes. Additionally, the $W1-W2$ color evolution also stays relatively constant; thus, we can assume that the effective temperature and bolometric correction of the source stays constant. Taking into account the fading since outburst peak, we find the peak luminosity of the outburst to be around $260$\,L$_\odot$, which is in accordance with other FU Ori sources \citep{Connelley2018, Zakri2022}.


\subsection{Spectroscopic Features}

\begin{figure*}
    \centering
        \includegraphics[width=\textwidth]{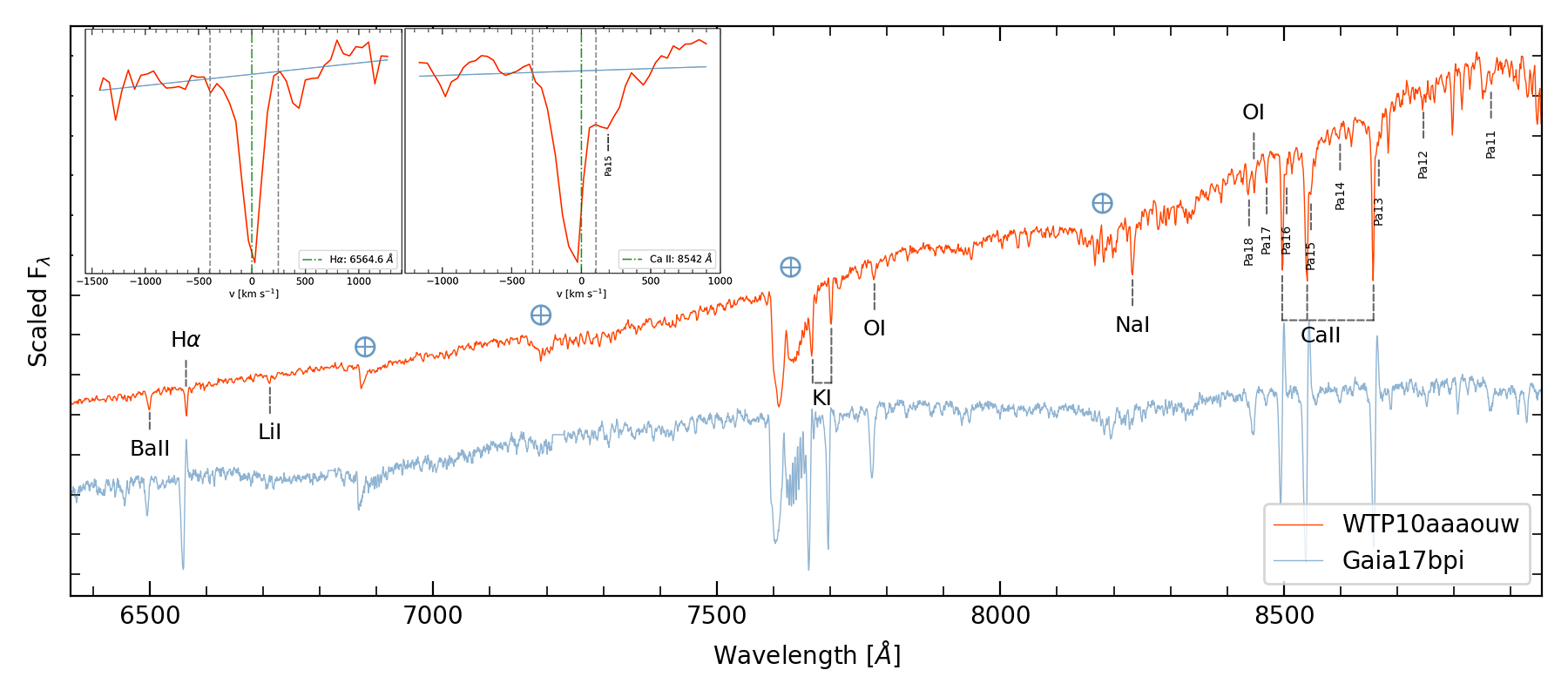}
    \caption{SOAR red optical spectrum of WTP\,10aaauow (orange) compared to that of the FU Ori outburst Gaia\,17bpi (blue; \citealt{Hillenbrand2018}). Regions with significant telluric contamination are marked with $\oplus$, while prominent lines are marked and indicated with dashed lines. The two insets show zoom-ins of H$\alpha$ and Ca\,II 8542\,\AA\ line regions, and are plotted in rest-frame velocity units. The green dash-dotted lines indicate rest frame (zero velocity) positions, the blue lines show linear fits of the surrounding continuum to highlight the absorption, and the dashed grey lines mark the terminal edges of the spectral lines. The Paschen\,15 line is shown in the Ca\,II 8542\,\AA\ inset, explaining the profile's extra red-shifted absorption region.}
    \label{fig:vs}
\end{figure*}


The optical spectrum of WTP\,10aaauow is shown in Figure \ref{fig:vs} in comparison to the recent FU Ori outburst source Gaia\,17bpi \citep{Hillenbrand2018}, which has a typical spectrum for the class, showing a mix of higher and lower excitation potential lines. The absorption aspects of the two spectra are similar in many respects.  We clearly detect in WTP\,10aaauow features of Ba\,II, Ca\,I and Fe\,I, together with a prominent Li\,I line at 6707\,\AA\ with an equivalent width of $W_\lambda = 0.48 $\,\AA\ -- a strong signature of youth. These signatures demonstrate that the optical spectrum of  WTP\,10aaauow is characteristic of GK-type stars as seen in other FU Ori outbursts. 

However, the wind indicators in  WTP\,10aaauow are weaker than those near the outburst epoch of Gaia\,17bpi, likely because our spectrum was obtained 7 years after the onset of the WTP\,10aaauow outburst and FU Ori winds can evolve quickly \citep{carvalho2023}. Nevertheless, the prominent wind lines of H$\alpha$, K\,I, O\,I, and the Ca\,II triplet, all show asymmetric profiles and terminal velocities -- as measured from the bluest edge of the absorption lines -- of several hundred km\,s$^{-1}$, indicating the presence of a strong wind. In particular, the H$\alpha$ and 8542\,\AA\ Ca\,II lines (highlighted in Figure \ref{fig:vs}) have terminal velocities approaching $\approx -400$\,km\,s$^{-1}$. 
We measure the equivalent width of the H$\alpha$ absorption to be $W_\lambda = 2.1$\,\AA, and $W_\lambda = 4.0 $\,\AA\ for the 8542\,\AA\ Ca\,II line. In addition, the upper Paschen lines of \ion{H}{1} are seen, which may arise at least partially in a wind.

\begin{figure*}
    \centering
        \includegraphics[width=0.93 \textwidth]{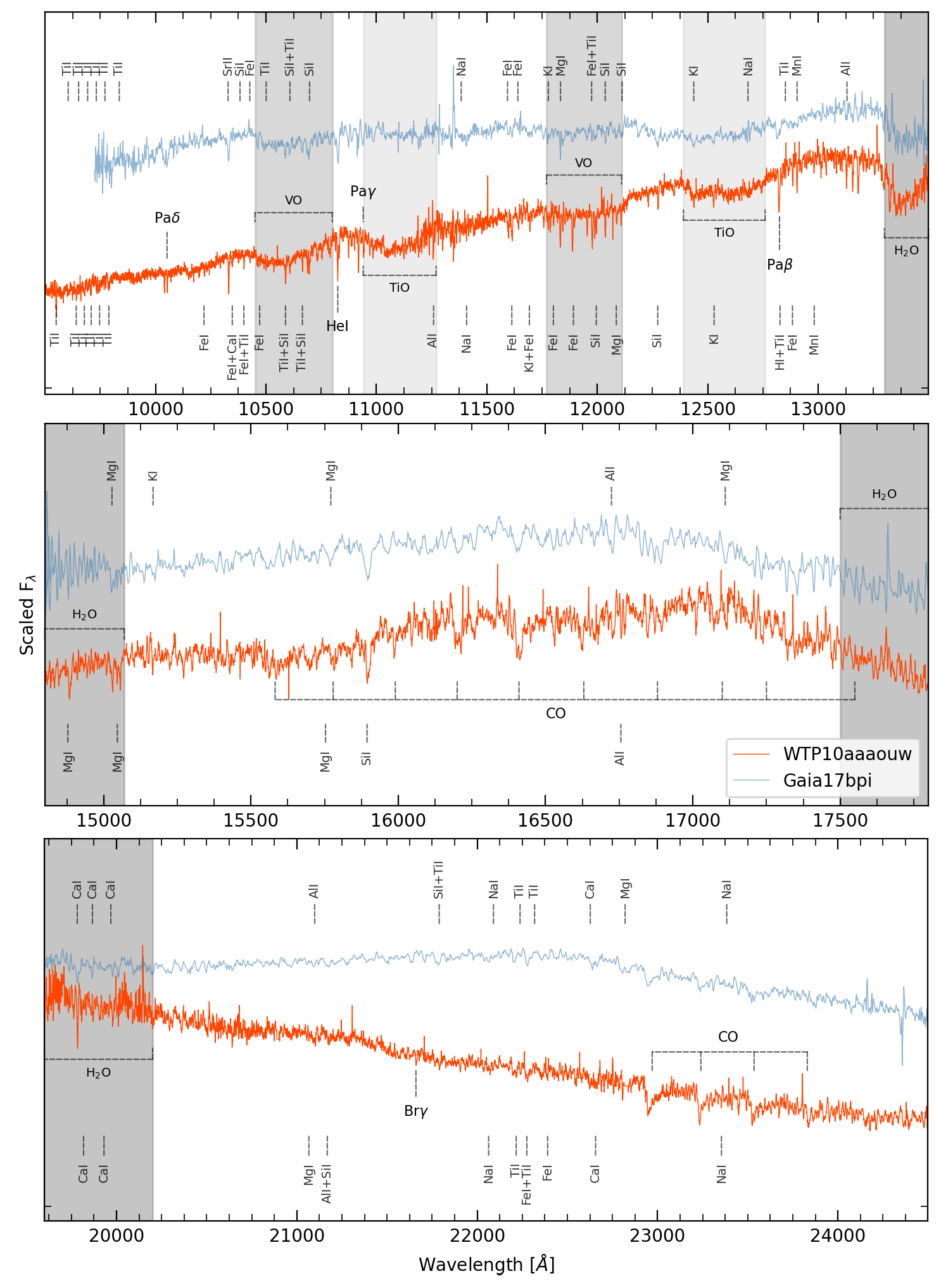}
    \caption{From top to bottom, the combined $YJ$, $H$, and $K$ band spectrum of WTP\,10aaauow (orange), compared to those of Gaia 17\,bpi (blue). Prominent lines are marked with their names and dashed lines; additional metal lines are approximately annotated with smaller markers. TiO, VO, and H$_2$O bands are marked with light, medium and dark gray regions, respectively.}
    \label{fig:ir}
\end{figure*}

The NIR spectrum of WTP\,10aaauow in Figure \ref{fig:ir},  shows classic signatures of FU Ori outbursts -- strong absorption $^{12}$CO(2,0) band-head in the 2.3\,$\mu$m region seen in late K, M stars, as well as the H$_2$O bands from $1.3-1.5\,\mu$m and $1.75-2.05\,\mu$m \citep{Connelley2018}. We detect absorption in Na\,I and Ca\,I lines in $K$-band, and in the Mg\,I and Si\,I lines in $H$-band. At shorter wavelengths, VO bands (centered at 1.06 and 1.19 $\mu$m), as well as TiO bands (centered at $1.11$ and $1.26\,\mu$m) are visible characteristic of late M-type spectra. It compares well to Gaia\,17bpi which is typical of the FU Ori class spectroscopically
and has the same mix of atomic and molecular features.

We measure the equivalent widths (following the prescription of \citealt{Messineo2021}) of the $^{12}$CO(2,0) line to be $W_\lambda = 25.85 $\,\AA, $W_\lambda = 1.65 $\,\AA\ for the 2.206 and 2.209 $\mu$m Na\,I lines and $W_\lambda = 2.17$\,\AA\ for the 2.263 and 2.266 $\mu$m Ca\,I lines. Comparing these values to the spectroscopic diagnostics presented in Figure 10 of \citet{Connelley2018} squarely places WTP\,10aaauow in the region occupied by FU Ori and FU Ori-like objects. The (CO, Na$+$Ca) equivalence width phase space value of WTP\,10aaauow, (25.85, 3.82), is very close to those of other FU Ori objects such as V1515\,Cyg (26.31, 3.79), V900\,Mon (27.39, 4.10), and FU Ori (28.69, 4.36), as well as those of FU Ori-like objects such as Parsamian\,21 (23.54, 3.94) and L1551\,IRS\,5 (28.77, 3.80). In terms of wind signatures, the He\,I 10830\,\AA\ transition is detected, indicating hot gas, and exhibits a blue-shifted center of absorption (by $\approx -100$\,km\,s$^{-1}$) similar to the optical Ca\,II triplet lines. While the lower \ion{H}{1} Paschen transitions are detected in the NIR region, the Br series is not clearly detected other than some weak Br$\gamma$.

\section{Summary and Discussion}
\label{sec:discussion}

\begin{table}
    \centering
    \begin{tabular}{|| c | c ||}
        \hline
        \hline
        R.A. (J2000) & 10:26:15.98 \\
        DEC. (J2000) & -58:20:37.80 \\
        Distance [kpc] & $4.0\pm0.4$ \\
        \hline
        $A\rm _V$ [mag] & 5-7 \\
        $T_{progenitor}$ [K] & $3600-4000$ \\
        \hline
        Optical Amplitude [mag] & $\gtrsim 5$ \\
        NIR Amplitude [mag] & $\gtrsim 5$ \\
        MIR Amplitude [mag] & $\approx 3$ ($\approx 1.2$) \\
        \hline
        Rise time [month] & $\approx 12$ ($\approx 60$) \\
        Rise rate [mag/month] & $\approx 0.25$ ($\approx 0.02$) \\
        Decay rate [mag/month] & $\approx 0.01$ \\
        \hline
        $L_{peak}$ [L$_\odot$] & $\approx$260 \\
        \hline
    \end{tabular}
    \caption{Summary information of WTP\,10aaauow. The rise time, rise rate, and decay rate are quoted from MIR bands. For the MIR amplitude, rise time, and rise rate, values from the steep rise between 2014 and 2015 are primarily shown with values from the slow rise between 2010 and 2014 in parenthesis.}
    \label{tab:cross-section}    
\end{table}

We have presented the discovery and follow-up observations of a large-amplitude outburst WTP\,10aaauow in the southern Galactic disk, located towards the RCW\,49 star forming region. 
We provide a quantitative summary of the properties of the source in Table \ref{tab:cross-section}.

Our discovery was enabled with the implementation of an image subtraction pipeline to search for MIR transients directly in NEOWISE data. Combined with archival observations from Spitzer, 2MASS, Gaia, and the DECam legacy surveys, we show that the source exhibited a $\gtrsim 5$\,mag outburst in the NIR bands (compared to archival measurements from 2MASS) peaking in 2015, followed by a slow decline. The MIR light curve shows a slow rise between 2010 and 2014 (at a rate of $\approx 0.02$\,mag\,month$^{-1}$
followed by a steep rise of $\approx 3$\,mag (at a rate of $\approx 0.25$\,mag\,month$^{-1}$) between 2014 and 2015. The observed shallow rise prior to the steeper brightening to peak exhibited by WTP\,10aaauow, was also documented in HBC 722 by \cite{miller2011}. The observed rapid rise timescale to peak is similar to that seen in other FU Ori outbursts \citep{Hillenbrand2018, Hillenbrand2021}. The final part of the brightening was also detected with the Gaia mission, exhibiting a $\gtrsim 5$\,mag rise in the optical bands, while simultaneously becoming bluer in both optical and MIR colors. Archival optical images from the DECam Galactic plane survey (taken after the outburst peak) clearly show extended nebulosity around the source. The source has subsequently exhibited a plateau/slow decline of $\approx 0.01$\,mag\,month$^{-1}$.

The spectroscopic characteristics of the source are strikingly similar to those of other FU Ori outbursts in the 0.3-2.5 $\mu$m region \citep[see e.g.][for in-depth discussion]{carvalho2023}. In the optical, we see a GK-type spectrum with asymmetric, complex line profiles having terminal velocities reaching $400$\,km\,s$^{-1}$, that likely arise from a wind, and strong Li\,I absorption. In the NIR, we observed a cooler M-type spectrum exhibiting a rich forest of atomic and molecular lines, suggestive of a mixed temperature spectrum. Quantitatively, we show the equivalent widths of the CO, Na\,I and Ca\,I lines in $K$-band are consistent with those seen in FU Ori stars \citep{Connelley2018}.

By examining the environment of the source to identify likely nearby young stars with infrared excesses, we find a large number ($\approx 176$) of nearby sources at distances consistent with that of the RCW\,49 region as well as the outburst itself (as measured by Gaia; $\approx 4$\,kpc). The estimated distance places WTP\,10aaauow in the outskirts of the RCW\,49 region. Fitting stellar templates to pre-outburst photometry, we invoke the presence of a two-component dust disk (at temperatures of $\approx 120$ and $\approx 700$\,K) to explain the observed infrared excess. The corresponding photospheric temperatures and foreground extinctions range between $3600 - 4000$\,K, and $5 - 7$\,mag respectively. In quiescence, the spectral slope in the infrared ($\alpha \approx 0.14$) is consistent with that of a Class I or flat spectrum YSO.

 To place WTP\,10aaauow in context, we first note that all properties of the outburst source seem typical for this small class of objects. As mentioned earlier, both the optical and infrared spectra have the expected absorption-dominated features arising from particular atomic and molecular lines. These lines indicate a mixed spectrum with a hotter photosphere towards the blue and a cooler photosphere towards the red. The source luminosity of $\approx 260$\,L$_\odot$ is also typical of the few hundred $L_\odot$ observed for FU Ori outbursts \citep{Connelley2018}. 

WTP\,10aaauow is rare among known members of the class, however, in that the photometric observations have managed to sample the object in its pre-outburst stage, during the outburst, and well into its outburst. 
Furthermore, there are well-sampled colors that show the source becoming bluer in both the optical and the infrared as it brightened. This is uncommon, if not unique. Overall, WTP\,10aaauow appears to be an excellent example of an FU Ori outburst underway. 
 WTP\,10aaauow thus adds to a rare class of FU Ori outbursts with well-sampled light curves in both the optical and MIR bands. The other examples are Gaia\,17bpi \citep{Hillenbrand2018} and Gaia\,18dvy \citep{Szegedi-Elek2020}. WTP\,10aaauow's exquisitely sampled Gaia light curve near the peak of the outburst clearly shows the outburst amplitude is larger in the optical bands compared to the MIR, consistent with expectations for a dramatically increasing accretion rate \citep[e.g.][]{Hillenbrand2022}. The multi-wavelength lightcurve of this source should be an excellent example for comparison to the outburst profiles and lightcurves now being generated by disk instability theorists, as in \cite{cleaver2023} and \cite{nayakshin2024b}.

In conclusion, we note that while optical surveys have succeeded at identifying YSOs undergoing accretion outbursts (e.g. Gaia 17bpi and Gaia 18dvy noted above), the long temporal baseline and MIR sensitivity of NEOWISE offers the best opportunity to systematically identify luminous FU Ori outbursts in the most obscured regions ($A_V \gtrsim 10$\,mag) of the Galactic plane, where their faint optical counterparts may be completely undetected or confused with other stellar outbursts. In particular, MIR searches provide unique sensitivity to outbursts in the earliest stages of star formation when the sources are still embedded and the rate of FU Ori outbursts are expected to be highest \citep{Bae2014}. Finally, the identification of WTP\,10aaauow as a bright MIR transient that saturates in WISE data, but yet overlooked thus far highlights the importance of applying systematic transient recovery techniques on the NEOWISE dataset to uncover the full population of recent FU Ori outbursts in the Galactic plane. In particular, we note the importance of image subtraction techniques in dense Galactic stellar fields where sources may be easily confused at the spatial resolution of WISE, as well as in the published point source catalogs.











\section*{Acknowledgements}
We thank the anonymous referee for a careful review of our work. We thank Ryan Lau for providing access to the LCO time for follow-up of the source. We thank J. J. Hermes, A. Meisner, and R. Simcoe for valuable comments. We thank A. Meisner for assistance with processing the \texttt{unwise} data. K. D. was supported by NASA through the NASA Hubble Fellowship grant \#HST-HF2-51477.001 awarded by the Space Telescope Science Institute, which is operated by the Association of Universities for Research in Astronomy, Inc., for NASA, under contract NAS5-26555. Based on observations obtained at the Southern Astrophysical Research (SOAR) telescope, which is a joint project of the Minist\'{e}rio da Ci\^{e}ncia, Tecnologia e Inova\c{c}\~{o}es (MCTI/LNA) do Brasil, the US National Science Foundation’s NOIRLab, the University of North Carolina at Chapel Hill (UNC), and Michigan State University (MSU). This paper includes data gathered with the 6.5-meter Magellan Telescopes located at Las Campanas Observatory, Chile. A part of the LCO follow-up observations is supported by the International Top Young-Fellow (ITYF) program of JAXA. We thank Mansi Kasliwal, Josh Bloom, and the SkyPortal team for assistance with using the data platform for this project.

\section*{Data Availability}

The photometry and spectra will be made available as machine readable files as part of the online publication.



\bibliographystyle{mnras}
\bibliography{example} 

\begin{thebibliography}{}
\makeatletter
\relax
\def\mn@urlcharsother{\let\do\@makeother \do\$\do\&\do\#\do\^\do\_\do\%\do\~}
\def\mn@doi{\begingroup\mn@urlcharsother \@ifnextchar [ {\mn@doi@}
  {\mn@doi@[]}}
\def\mn@doi@[#1]#2{\def\@tempa{#1}\ifx\@tempa\@empty \href
  {http://dx.doi.org/#2} {doi:#2}\else \href {http://dx.doi.org/#2} {#1}\fi
  \endgroup}
\def\mn@eprint#1#2{\mn@eprint@#1:#2::\@nil}
\def\mn@eprint@arXiv#1{\href {http://arxiv.org/abs/#1} {{\tt arXiv:#1}}}
\def\mn@eprint@dblp#1{\href {http://dblp.uni-trier.de/rec/bibtex/#1.xml}
  {dblp:#1}}
\def\mn@eprint@#1:#2:#3:#4\@nil{\def\@tempa {#1}\def\@tempb {#2}\def\@tempc
  {#3}\ifx \@tempc \@empty \let \@tempc \@tempb \let \@tempb \@tempa \fi \ifx
  \@tempb \@empty \def\@tempb {arXiv}\fi \@ifundefined
  {mn@eprint@\@tempb}{\@tempb:\@tempc}{\expandafter \expandafter \csname
  mn@eprint@\@tempb\endcsname \expandafter{\@tempc}}}

\bibitem[\protect\citeauthoryear{{Armitage}}{{Armitage}}{2015}]{armitage2015}
{Armitage} P.~J.,  2015, \mn@doi [arXiv e-prints] {10.48550/arXiv.1509.06382},
  \href {https://ui.adsabs.harvard.edu/abs/2015arXiv150906382A} {p.
  arXiv:1509.06382}

\bibitem[\protect\citeauthoryear{{Armitage}}{{Armitage}}{2019}]{Armitage2019}
{Armitage} P.~J.,  2019, in {Audard} M.,  {Meyer} M.~R.,   {Alibert} Y.,  eds,
  Saas-Fee Advanced Course Vol. 45, Saas-Fee Advanced Course. p.~1,
  \mn@doi{10.1007/978-3-662-58687-7_1}

\bibitem[\protect\citeauthoryear{{Ashraf}, {Jose}, {Lee}, {Pe{\~n}a},
  {Herczeg}, {Liu}, {Johnstone}  \& {Lee}}{{Ashraf} et~al.}{2023}]{ashraf2023}
{Ashraf} M.,  {Jose} J.,  {Lee} H.-G.,  {Pe{\~n}a} C.~C.,  {Herczeg} G.,  {Liu}
  H.,  {Johnstone} D.,   {Lee} J.-E.,  2023, \mn@doi [\mnras]
  {10.1093/mnras/stad3900}, \href
  {https://ui.adsabs.harvard.edu/abs/2023MNRAS.tmp.3745A} {}

\bibitem[\protect\citeauthoryear{{Audard} et~al.,}{{Audard}
  et~al.}{2014}]{auduard2014}
{Audard} M.,  et~al., 2014, in {Beuther} H.,  {Klessen} R.~S.,  {Dullemond}
  C.~P.,   {Henning} T.,  eds, Protostars and Planets VI. pp 387--410
  (\mn@eprint {arXiv} {1401.3368}),
  \mn@doi{10.2458/azu_uapress_9780816531240-ch017}

\bibitem[\protect\citeauthoryear{{Bae}, {Hartmann}, {Zhu}  \& {Nelson}}{{Bae}
  et~al.}{2014}]{Bae2014}
{Bae} J.,  {Hartmann} L.,  {Zhu} Z.,   {Nelson} R.~P.,  2014, \mn@doi [\apj]
  {10.1088/0004-637X/795/1/61}, \href
  {https://ui.adsabs.harvard.edu/abs/2014ApJ...795...61B} {795, 61}

\bibitem[\protect\citeauthoryear{{Benjamin} et~al.,}{{Benjamin}
  et~al.}{2003}]{Benjamin2003}
{Benjamin} R.~A.,  et~al., 2003, \mn@doi [\pasp] {10.1086/376696}, \href
  {https://ui.adsabs.harvard.edu/abs/2003PASP..115..953B} {115, 953}

\bibitem[\protect\citeauthoryear{{Brown} et~al.,}{{Brown}
  et~al.}{2013}]{Brown2013}
{Brown} T.~M.,  et~al., 2013, \mn@doi [\pasp] {10.1086/673168}, \href
  {https://ui.adsabs.harvard.edu/abs/2013PASP..125.1031B} {125, 1031}

\bibitem[\protect\citeauthoryear{{Carvalho} \& {Hillenbrand}}{{Carvalho} \&
  {Hillenbrand}}{2022}]{Carvalho2022}
{Carvalho} A.~S.,  {Hillenbrand} L.~A.,  2022, \mn@doi [\apj]
  {10.3847/1538-4357/ac9d8e}, \href
  {https://ui.adsabs.harvard.edu/abs/2022ApJ...940..156C} {940, 156}

\bibitem[\protect\citeauthoryear{{Carvalho}, {Hillenbrand}  \&
  {Seebeck}}{{Carvalho} et~al.}{2023}]{carvalho2023}
{Carvalho} A.~S.,  {Hillenbrand} L.~A.,   {Seebeck} J.,  2023, \mn@doi [arXiv
  e-prints] {10.48550/arXiv.2310.02465}, \href
  {https://ui.adsabs.harvard.edu/abs/2023arXiv231002465C} {p. arXiv:2310.02465}

\bibitem[\protect\citeauthoryear{{Churchwell} et~al.,}{{Churchwell}
  et~al.}{2004}]{Churchwell2004}
{Churchwell} E.,  et~al., 2004, \mn@doi [\apjs] {10.1086/422504}, \href
  {https://ui.adsabs.harvard.edu/abs/2004ApJS..154..322C} {154, 322}

\bibitem[\protect\citeauthoryear{{Churchwell} et~al.,}{{Churchwell}
  et~al.}{2009}]{Churchwell2009}
{Churchwell} E.,  et~al., 2009, \mn@doi [\pasp] {10.1086/597811}, \href
  {https://ui.adsabs.harvard.edu/abs/2009PASP..121..213C} {121, 213}

\bibitem[\protect\citeauthoryear{{Clarke}, {Lodato}, {Melnikov}  \&
  {Ibrahimov}}{{Clarke} et~al.}{2005}]{clarke2005}
{Clarke} C.,  {Lodato} G.,  {Melnikov} S.~Y.,   {Ibrahimov} M.~A.,  2005,
  \mn@doi [\mnras] {10.1111/j.1365-2966.2005.09231.x}, \href
  {https://ui.adsabs.harvard.edu/abs/2005MNRAS.361..942C} {361, 942}

\bibitem[\protect\citeauthoryear{{Cleaver}, {Hartmann}  \& {Bae}}{{Cleaver}
  et~al.}{2023}]{cleaver2023}
{Cleaver} J.,  {Hartmann} L.,   {Bae} J.,  2023, \mn@doi [\mnras]
  {10.1093/mnras/stad1784}, \href
  {https://ui.adsabs.harvard.edu/abs/2023MNRAS.523.5522C} {523, 5522}

\bibitem[\protect\citeauthoryear{{Clemens}, {Crain}  \& {Anderson}}{{Clemens}
  et~al.}{2004}]{Clemens2004}
{Clemens} J.~C.,  {Crain} J.~A.,   {Anderson} R.,  2004, in {Moorwood} A.
  F.~M.,  {Iye} M.,  eds,  Society of Photo-Optical Instrumentation Engineers
  (SPIE) Conference Series Vol. 5492, Ground-based Instrumentation for
  Astronomy. pp 331--340, \mn@doi{10.1117/12.550069}

\bibitem[\protect\citeauthoryear{{Connelley} \& {Reipurth}}{{Connelley} \&
  {Reipurth}}{2018}]{Connelley2018}
{Connelley} M.~S.,  {Reipurth} B.,  2018, \mn@doi [\apj]
  {10.3847/1538-4357/aaba7b}, \href
  {https://ui.adsabs.harvard.edu/abs/2018ApJ...861..145C} {861, 145}

\bibitem[\protect\citeauthoryear{{Contreras Pe{\~n}a} et~al.,}{{Contreras
  Pe{\~n}a} et~al.}{2017}]{Contreras-Pena2017}
{Contreras Pe{\~n}a} C.,  et~al., 2017, \mn@doi [\mnras]
  {10.1093/mnras/stw2801}, \href
  {https://ui.adsabs.harvard.edu/abs/2017MNRAS.465.3011C} {465, 3011}

\bibitem[\protect\citeauthoryear{{Contreras Pe{\~n}a}, {Naylor}  \&
  {Morrell}}{{Contreras Pe{\~n}a} et~al.}{2019}]{Contreras-Pena2019}
{Contreras Pe{\~n}a} C.,  {Naylor} T.,   {Morrell} S.,  2019, \mn@doi [\mnras]
  {10.1093/mnras/stz1019}, \href
  {https://ui.adsabs.harvard.edu/abs/2019MNRAS.486.4590C} {486, 4590}

\bibitem[\protect\citeauthoryear{{Cutri} et~al.,}{{Cutri}
  et~al.}{2003}]{Cutri2003}
{Cutri} R.~M.,  et~al., 2003, {2MASS All Sky Catalog of point sources.}

\bibitem[\protect\citeauthoryear{{De} et~al.,}{{De} et~al.}{2020a}]{De2019}
{De} K.,  et~al., 2020a, \mn@doi [\pasp] {10.1088/1538-3873/ab6069}, \href
  {https://ui.adsabs.harvard.edu/abs/2020PASP..132b5001D} {132, 025001}

\bibitem[\protect\citeauthoryear{{De} et~al.,}{{De} et~al.}{2020b}]{De2020}
{De} K.,  et~al., 2020b, \mn@doi [\apj] {10.3847/1538-4357/abb45c}, \href
  {https://ui.adsabs.harvard.edu/abs/2020ApJ...905...58D} {905, 58}

\bibitem[\protect\citeauthoryear{{Dey} et~al.,}{{Dey} et~al.}{2019}]{Dey2019}
{Dey} A.,  et~al., 2019, \mn@doi [\aj] {10.3847/1538-3881/ab089d}, \href
  {https://ui.adsabs.harvard.edu/abs/2019AJ....157..168D} {157, 168}

\bibitem[\protect\citeauthoryear{{Drew} et~al.,}{{Drew} et~al.}{2014}]{vphas}
{Drew} J.~E.,  et~al., 2014, \mn@doi [\mnras] {10.1093/mnras/stu394}, \href
  {https://ui.adsabs.harvard.edu/abs/2014MNRAS.440.2036D} {440, 2036}

\bibitem[\protect\citeauthoryear{{Fischer}, {Hillenbrand}, {Herczeg},
  {Johnstone}, {Kospal}  \& {Dunham}}{{Fischer} et~al.}{2023}]{fischer2023}
{Fischer} W.~J.,  {Hillenbrand} L.~A.,  {Herczeg} G.~J.,  {Johnstone} D.,
  {Kospal} A.,   {Dunham} M.~M.,  2023, in {Inutsuka} S.,  {Aikawa} Y.,  {Muto}
  T.,  {Tomida} K.,   {Tamura} M.,  eds,  Astronomical Society of the Pacific
  Conference Series Vol. 534, Protostars and Planets VII. p.~355 (\mn@eprint
  {arXiv} {2203.11257}), \mn@doi{10.48550/arXiv.2203.11257}

\bibitem[\protect\citeauthoryear{{Fouesneau}, {Andrae}, {Dharmawardena},
  {Rybizki}, {Bailer-Jones}  \& {Demleitner}}{{Fouesneau}
  et~al.}{2022}]{fouesneau2022}
{Fouesneau} M.,  {Andrae} R.,  {Dharmawardena} T.,  {Rybizki} J.,
  {Bailer-Jones} C.~A.~L.,   {Demleitner} M.,  2022, \mn@doi [\aap]
  {10.1051/0004-6361/202141828}, \href
  {https://ui.adsabs.harvard.edu/abs/2022A&A...662A.125F} {662, A125}

\bibitem[\protect\citeauthoryear{{Gaia Collaboration} et~al.,}{{Gaia
  Collaboration} et~al.}{2022}]{GaiaCollaboration2022}
{Gaia Collaboration} et~al., 2022, \mn@doi [arXiv e-prints]
  {10.48550/arXiv.2208.00211}, \href
  {https://ui.adsabs.harvard.edu/abs/2022arXiv220800211G} {p. arXiv:2208.00211}

\bibitem[\protect\citeauthoryear{{Greene}, {Wilking}, {Andre}, {Young}  \&
  {Lada}}{{Greene} et~al.}{1994}]{greene1994}
{Greene} T.~P.,  {Wilking} B.~A.,  {Andre} P.,  {Young} E.~T.,   {Lada} C.~J.,
  1994, \mn@doi [\apj] {10.1086/174763}, \href
  {https://ui.adsabs.harvard.edu/abs/1994ApJ...434..614G} {434, 614}

\bibitem[\protect\citeauthoryear{{Guo} et~al.,}{{Guo} et~al.}{2021}]{Guo2021}
{Guo} Z.,  et~al., 2021, \mn@doi [\mnras] {10.1093/mnras/stab882}, \href
  {https://ui.adsabs.harvard.edu/abs/2021MNRAS.504..830G} {504, 830}

\bibitem[\protect\citeauthoryear{{Guo} et~al.,}{{Guo} et~al.}{2022}]{Guo2022}
{Guo} Z.,  et~al., 2022, \mn@doi [\mnras] {10.1093/mnras/stac768}, \href
  {https://ui.adsabs.harvard.edu/abs/2022MNRAS.513.1015G} {513, 1015}

\bibitem[\protect\citeauthoryear{{Gutermuth}, {Megeath}, {Myers}, {Allen},
  {Pipher}  \& {Fazio}}{{Gutermuth} et~al.}{2009}]{Gutermuth2009}
{Gutermuth} R.~A.,  {Megeath} S.~T.,  {Myers} P.~C.,  {Allen} L.~E.,  {Pipher}
  J.~L.,   {Fazio} G.~G.,  2009, \mn@doi [\apjs] {10.1088/0067-0049/184/1/18},
  \href {https://ui.adsabs.harvard.edu/abs/2009ApJS..184...18G} {184, 18}

\bibitem[\protect\citeauthoryear{{Hartmann} \& {Kenyon}}{{Hartmann} \&
  {Kenyon}}{1996}]{hartmann1996}
{Hartmann} L.,  {Kenyon} S.~J.,  1996, \mn@doi [\araa]
  {10.1146/annurev.astro.34.1.207}, \href
  {https://ui.adsabs.harvard.edu/abs/1996ARA&A..34..207H} {34, 207}

\bibitem[\protect\citeauthoryear{{Hauschildt}, {Allard}, {Ferguson}, {Baron}
  \& {Alexander}}{{Hauschildt} et~al.}{1999}]{Hauschildt1999}
{Hauschildt} P.~H.,  {Allard} F.,  {Ferguson} J.,  {Baron} E.,   {Alexander}
  D.~R.,  1999, \mn@doi [\apj] {10.1086/307954}, \href
  {https://ui.adsabs.harvard.edu/abs/1999ApJ...525..871H} {525, 871}

\bibitem[\protect\citeauthoryear{{Herbig}}{{Herbig}}{1977}]{Herbig1977}
{Herbig} G.~H.,  1977, \mn@doi [\apj] {10.1086/155615}, \href
  {https://ui.adsabs.harvard.edu/abs/1977ApJ...217..693H} {217, 693}

\bibitem[\protect\citeauthoryear{{Hillenbrand} \& {Rodriguez}}{{Hillenbrand} \&
  {Rodriguez}}{2022}]{Hillenbrand2022}
{Hillenbrand} L.~A.,  {Rodriguez} A.~C.,  2022, \mn@doi [Research Notes of the
  American Astronomical Society] {10.3847/2515-5172/ac4807}, \href
  {https://ui.adsabs.harvard.edu/abs/2022RNAAS...6....6H} {6, 6}

\bibitem[\protect\citeauthoryear{{Hillenbrand} et~al.,}{{Hillenbrand}
  et~al.}{2018}]{Hillenbrand2018}
{Hillenbrand} L.~A.,  et~al., 2018, \mn@doi [\apj] {10.3847/1538-4357/aaf414},
  \href {https://ui.adsabs.harvard.edu/abs/2018ApJ...869..146H} {869, 146}

\bibitem[\protect\citeauthoryear{{Hillenbrand} et~al.,}{{Hillenbrand}
  et~al.}{2021}]{Hillenbrand2021}
{Hillenbrand} L.~A.,  et~al., 2021, \mn@doi [\aj] {10.3847/1538-3881/abe406},
  \href {https://ui.adsabs.harvard.edu/abs/2021AJ....161..220H} {161, 220}

\bibitem[\protect\citeauthoryear{{Hillenbrand}, {Carvalho}, {van Roestel}  \&
  {De}}{{Hillenbrand} et~al.}{2023}]{hillenbrand2023}
{Hillenbrand} L.~A.,  {Carvalho} A.,  {van Roestel} J.,   {De} K.,  2023,
  \mn@doi [\apjl] {10.3847/2041-8213/ad0be0}, \href
  {https://ui.adsabs.harvard.edu/abs/2023ApJ...958L..27H} {958, L27}

\bibitem[\protect\citeauthoryear{{Hodgkin} et~al.,}{{Hodgkin}
  et~al.}{2021}]{gaia_alerts}
{Hodgkin} S.~T.,  et~al., 2021, \mn@doi [\aap] {10.1051/0004-6361/202140735},
  \href {https://ui.adsabs.harvard.edu/abs/2021A&A...652A..76H} {652, A76}

\bibitem[\protect\citeauthoryear{{Hosokawa}, {Offner}  \&
  {Krumholz}}{{Hosokawa} et~al.}{2011}]{hosokawa2011}
{Hosokawa} T.,  {Offner} S. S.~R.,   {Krumholz} M.~R.,  2011, \mn@doi [\apj]
  {10.1088/0004-637X/738/2/140}, \href
  {https://ui.adsabs.harvard.edu/abs/2011ApJ...738..140H} {738, 140}

\bibitem[\protect\citeauthoryear{{Kenyon}, {Hartmann}  \& {Hewett}}{{Kenyon}
  et~al.}{1988}]{Kenyon1988}
{Kenyon} S.~J.,  {Hartmann} L.,   {Hewett} R.,  1988, \mn@doi [\apj]
  {10.1086/165999}, \href
  {https://ui.adsabs.harvard.edu/abs/1988ApJ...325..231K} {325, 231}

\bibitem[\protect\citeauthoryear{{Kochanek} et~al.,}{{Kochanek}
  et~al.}{2017}]{Kochanek2017}
{Kochanek} C.~S.,  et~al., 2017, \mn@doi [\pasp] {10.1088/1538-3873/aa80d9},
  \href {https://ui.adsabs.harvard.edu/abs/2017PASP..129j4502K} {129, 104502}

\bibitem[\protect\citeauthoryear{{Kuhn}, {de Souza}, {Krone-Martins},
  {Castro-Ginard}, {Ishida}, {Povich}, {Hillenbrand}  \& {COIN
  Collaboration}}{{Kuhn} et~al.}{2021}]{kuhn2021}
{Kuhn} M.~A.,  {de Souza} R.~S.,  {Krone-Martins} A.,  {Castro-Ginard} A.,
  {Ishida} E. E.~O.,  {Povich} M.~S.,  {Hillenbrand} L.~A.,   {COIN
  Collaboration} 2021, \mn@doi [\apjs] {10.3847/1538-4365/abe465}, \href
  {https://ui.adsabs.harvard.edu/abs/2021ApJS..254...33K} {254, 33}

\bibitem[\protect\citeauthoryear{{Lang}}{{Lang}}{2014}]{Lang2014}
{Lang} D.,  2014, \mn@doi [\aj] {10.1088/0004-6256/147/5/108}, \href
  {https://ui.adsabs.harvard.edu/abs/2014AJ....147..108L} {147, 108}

\bibitem[\protect\citeauthoryear{{Loh}, {Biel}, {Davis}, {Laporte}, {Loh}  \&
  {Verhanovitz}}{{Loh} et~al.}{2012}]{Loh2012}
{Loh} E.~D.,  {Biel} J.~D.,  {Davis} M.~W.,  {Laporte} R.,  {Loh} O.~Y.,
  {Verhanovitz} N.~J.,  2012, \mn@doi [\pasp] {10.1086/665597}, \href
  {https://ui.adsabs.harvard.edu/abs/2012PASP..124..343L} {124, 343}

\bibitem[\protect\citeauthoryear{{Mainzer} et~al.,}{{Mainzer}
  et~al.}{2014}]{Mainzer2014}
{Mainzer} A.,  et~al., 2014, \mn@doi [\apj] {10.1088/0004-637X/792/1/30}, \href
  {https://ui.adsabs.harvard.edu/abs/2014ApJ...792...30M} {792, 30}

\bibitem[\protect\citeauthoryear{{Meisner}, {Lang}  \& {Schlegel}}{{Meisner}
  et~al.}{2018}]{Meisner2018}
{Meisner} A.~M.,  {Lang} D.,   {Schlegel} D.~J.,  2018, \mn@doi [\aj]
  {10.3847/1538-3881/aacbcd}, \href
  {https://ui.adsabs.harvard.edu/abs/2018AJ....156...69M} {156, 69}

\bibitem[\protect\citeauthoryear{{Messineo}, {Figer}, {Kudritzki}, {Zhu},
  {Menten}, {Ivanov}  \& {Chen}}{{Messineo} et~al.}{2021}]{Messineo2021}
{Messineo} M.,  {Figer} D.~F.,  {Kudritzki} R.-P.,  {Zhu} Q.,  {Menten} K.~M.,
  {Ivanov} V.~D.,   {Chen} C. H.~R.,  2021, \mn@doi [\aj]
  {10.3847/1538-3881/ac116b}, \href
  {https://ui.adsabs.harvard.edu/abs/2021AJ....162..187M} {162, 187}

\bibitem[\protect\citeauthoryear{{Miller} et~al.,}{{Miller}
  et~al.}{2011}]{miller2011}
{Miller} A.~A.,  et~al., 2011, \mn@doi [\apj] {10.1088/0004-637X/730/2/80},
  \href {https://ui.adsabs.harvard.edu/abs/2011ApJ...730...80M} {730, 80}

\bibitem[\protect\citeauthoryear{{Nagy} et~al.,}{{Nagy}
  et~al.}{2023}]{nagy2023}
{Nagy} Z.,  et~al., 2023, \mn@doi [\mnras] {10.1093/mnras/stad2019}, \href
  {https://ui.adsabs.harvard.edu/abs/2023MNRAS.524.3344N} {524, 3344}

\bibitem[\protect\citeauthoryear{{Nayakshin}, {Cruz Saenz de Miera}, {Kospal},
  {Calovic}, {Eisloffel}  \& {Lin}}{{Nayakshin} et~al.}{2024}]{nayakshin2024b}
{Nayakshin} S.,  {Cruz Saenz de Miera} F.,  {Kospal} A.,  {Calovic} A.,
  {Eisloffel} J.,   {Lin} D. N.~C.,  2024, \mn@doi [arXiv e-prints]
  {10.48550/arXiv.2403.04439}, \href
  {https://ui.adsabs.harvard.edu/abs/2024arXiv240304439N} {p. arXiv:2403.04439}

\bibitem[\protect\citeauthoryear{{Park} et~al.,}{{Park}
  et~al.}{2021}]{Park2021}
{Park} W.,  et~al., 2021, \mn@doi [\apj] {10.3847/1538-4357/ac1745}, \href
  {https://ui.adsabs.harvard.edu/abs/2021ApJ...920..132P} {920, 132}

\bibitem[\protect\citeauthoryear{{Pringle}}{{Pringle}}{1981}]{pringle1981}
{Pringle} J.~E.,  1981, \mn@doi [\araa] {10.1146/annurev.aa.19.090181.001033},
  \href {https://ui.adsabs.harvard.edu/abs/1981ARA&A..19..137P} {19, 137}

\bibitem[\protect\citeauthoryear{{Prochaska} et~al.,}{{Prochaska}
  et~al.}{2020}]{Prochaska2020}
{Prochaska} J.,  et~al., 2020, \mn@doi [The Journal of Open Source Software]
  {10.21105/joss.02308}, \href
  {https://ui.adsabs.harvard.edu/abs/2020JOSS....5.2308P} {5, 2308}

\bibitem[\protect\citeauthoryear{{Rodriguez} \& {Hillenbrand}}{{Rodriguez} \&
  {Hillenbrand}}{2022}]{Rodriguez2022}
{Rodriguez} A.~C.,  {Hillenbrand} L.~A.,  2022, \mn@doi [\apj]
  {10.3847/1538-4357/ac496b}, \href
  {https://ui.adsabs.harvard.edu/abs/2022ApJ...927..144R} {927, 144}

\bibitem[\protect\citeauthoryear{{Schlafly} \& {Finkbeiner}}{{Schlafly} \&
  {Finkbeiner}}{2011}]{Schlafly2011}
{Schlafly} E.~F.,  {Finkbeiner} D.~P.,  2011, \mn@doi [\apj]
  {10.1088/0004-637X/737/2/103}, \href
  {https://ui.adsabs.harvard.edu/abs/2011ApJ...737..103S} {737, 103}

\bibitem[\protect\citeauthoryear{{Schlafly} et~al.,}{{Schlafly}
  et~al.}{2018}]{Schlafly2018}
{Schlafly} E.~F.,  et~al., 2018, \mn@doi [\apjs] {10.3847/1538-4365/aaa3e2},
  \href {https://ui.adsabs.harvard.edu/abs/2018ApJS..234...39S} {234, 39}

\bibitem[\protect\citeauthoryear{{Shappee} et~al.,}{{Shappee}
  et~al.}{2014}]{Shappee2014}
{Shappee} B.~J.,  et~al., 2014, \mn@doi [\apj] {10.1088/0004-637X/788/1/48},
  \href {https://ui.adsabs.harvard.edu/abs/2014ApJ...788...48S} {788, 48}

\bibitem[\protect\citeauthoryear{{Simcoe} et~al.,}{{Simcoe}
  et~al.}{2013}]{Simcoe2013}
{Simcoe} R.~A.,  et~al., 2013, \mn@doi [\pasp] {10.1086/670241}, \href
  {https://ui.adsabs.harvard.edu/abs/2013PASP..125..270S} {125, 270}

\bibitem[\protect\citeauthoryear{{Skliarevskii} \& {Vorobyov}}{{Skliarevskii}
  \& {Vorobyov}}{2024}]{skliarevskii2024}
{Skliarevskii} A.~M.,  {Vorobyov} E.~I.,  2024, \mn@doi [Astronomy Reports]
  {10.1134/S1063772923120107}, \href
  {https://ui.adsabs.harvard.edu/abs/2024ARep...67.1401S} {67, 1401}

\bibitem[\protect\citeauthoryear{{Skrutskie} et~al.,}{{Skrutskie}
  et~al.}{2006}]{Skrutskie2006}
{Skrutskie} M.~F.,  et~al., 2006, \mn@doi [\aj] {10.1086/498708}, \href
  {https://ui.adsabs.harvard.edu/abs/2006AJ....131.1163S} {131, 1163}

\bibitem[\protect\citeauthoryear{{Spitzer Science Center (SSC)} \& {Infrared
  Science Archive (IRSA)}}{{Spitzer Science Center (SSC)} \& {Infrared Science
  Archive (IRSA)}}{2021}]{Spitzer2021}
{Spitzer Science Center (SSC)} {Infrared Science Archive (IRSA)} 2021, VizieR
  Online Data Catalog, \href
  {https://ui.adsabs.harvard.edu/abs/2021yCat.2368....0S} {p. II/368}

\bibitem[\protect\citeauthoryear{{Szegedi-Elek} et~al.,}{{Szegedi-Elek}
  et~al.}{2020}]{Szegedi-Elek2020}
{Szegedi-Elek} E.,  et~al., 2020, \mn@doi [\apj] {10.3847/1538-4357/aba129},
  \href {https://ui.adsabs.harvard.edu/abs/2020ApJ...899..130S} {899, 130}

\bibitem[\protect\citeauthoryear{{Turner}, {Fromang}, {Gammie}, {Klahr},
  {Lesur}, {Wardle}  \& {Bai}}{{Turner} et~al.}{2014}]{turner2014}
{Turner} N.~J.,  {Fromang} S.,  {Gammie} C.,  {Klahr} H.,  {Lesur} G.,
  {Wardle} M.,   {Bai} X.~N.,  2014, in {Beuther} H.,  {Klessen} R.~S.,
  {Dullemond} C.~P.,   {Henning} T.,  eds, Protostars and Planets VI. pp
  411--432 (\mn@eprint {arXiv} {1401.7306}),
  \mn@doi{10.2458/azu_uapress_9780816531240-ch018}

\bibitem[\protect\citeauthoryear{{Vacca}, {Cushing}  \& {Rayner}}{{Vacca}
  et~al.}{2003}]{Vacca2003}
{Vacca} W.~D.,  {Cushing} M.~C.,   {Rayner} J.~T.,  2003, \mn@doi [\pasp]
  {10.1086/346193}, \href
  {https://ui.adsabs.harvard.edu/abs/2003PASP..115..389V} {115, 389}

\bibitem[\protect\citeauthoryear{{Vioque}, {Oudmaijer}, {Schreiner},
  {Mendigut{\'\i}a}, {Baines}, {Mowlavi}  \&
  {P{\'e}rez-Mart{\'\i}nez}}{{Vioque} et~al.}{2020}]{vioque2020}
{Vioque} M.,  {Oudmaijer} R.~D.,  {Schreiner} M.,  {Mendigut{\'\i}a} I.,
  {Baines} D.,  {Mowlavi} N.,   {P{\'e}rez-Mart{\'\i}nez} R.,  2020, \mn@doi
  [\aap] {10.1051/0004-6361/202037731}, \href
  {https://ui.adsabs.harvard.edu/abs/2020A&A...638A..21V} {638, A21}

\bibitem[\protect\citeauthoryear{{Vorobyov}, {Elbakyan}, {Liu}  \&
  {Takami}}{{Vorobyov} et~al.}{2021}]{vorobyov2021}
{Vorobyov} E.~I.,  {Elbakyan} V.~G.,  {Liu} H.~B.,   {Takami} M.,  2021,
  \mn@doi [\aap] {10.1051/0004-6361/202039391}, \href
  {https://ui.adsabs.harvard.edu/abs/2021A&A...647A..44V} {647, A44}

\bibitem[\protect\citeauthoryear{{Welty}, {Strom}, {Edwards}, {Kenyon}  \&
  {Hartmann}}{{Welty} et~al.}{1992}]{Welty1992}
{Welty} A.~D.,  {Strom} S.~E.,  {Edwards} S.,  {Kenyon} S.~J.,   {Hartmann}
  L.~W.,  1992, \mn@doi [\apj] {10.1086/171785}, \href
  {https://ui.adsabs.harvard.edu/abs/1992ApJ...397..260W} {397, 260}

\bibitem[\protect\citeauthoryear{{Werner} et~al.,}{{Werner}
  et~al.}{2004}]{Werner2004}
{Werner} M.~W.,  et~al., 2004, \mn@doi [\apjs] {10.1086/422992}, \href
  {https://ui.adsabs.harvard.edu/abs/2004ApJS..154....1W} {154, 1}

\bibitem[\protect\citeauthoryear{{Whitney} et~al.,}{{Whitney}
  et~al.}{2004}]{Whitney2004}
{Whitney} B.~A.,  et~al., 2004, \mn@doi [\apjs] {10.1086/422557}, \href
  {https://ui.adsabs.harvard.edu/abs/2004ApJS..154..315W} {154, 315}

\bibitem[\protect\citeauthoryear{{Wright} et~al.,}{{Wright}
  et~al.}{2010}]{Wright2010}
{Wright} E.~L.,  et~al., 2010, \mn@doi [\aj] {10.1088/0004-6256/140/6/1868},
  \href {https://ui.adsabs.harvard.edu/abs/2010AJ....140.1868W} {140, 1868}

\bibitem[\protect\citeauthoryear{{Zackay}, {Ofek}  \& {Gal-Yam}}{{Zackay}
  et~al.}{2016}]{Zackay2016}
{Zackay} B.,  {Ofek} E.~O.,   {Gal-Yam} A.,  2016, \mn@doi [\apj]
  {10.3847/0004-637X/830/1/27}, \href
  {https://ui.adsabs.harvard.edu/abs/2016ApJ...830...27Z} {830, 27}

\bibitem[\protect\citeauthoryear{{Zakri} et~al.,}{{Zakri}
  et~al.}{2022}]{Zakri2022}
{Zakri} W.,  et~al., 2022, \mn@doi [\apjl] {10.3847/2041-8213/ac46ae}, \href
  {https://ui.adsabs.harvard.edu/abs/2022ApJ...924L..23Z} {924, L23}

\bibitem[\protect\citeauthoryear{{van der Walt}, {Crellin-Quick}  \&
  {Bloom}}{{van der Walt} et~al.}{2019}]{vanderWalt2019}
{van der Walt} S.,  {Crellin-Quick} A.,   {Bloom} J.,  2019, \mn@doi [The
  Journal of Open Source Software] {10.21105/joss.01247}, \href
  {https://ui.adsabs.harvard.edu/abs/2019JOSS....4.1247V} {4, 1247}

\makeatother
\end{thebibliography}




\appendix

\section{Photometry Measurements of WTP\,10aaauow}
\label{sec:appendix}

\begin{table}
    \centering
    \begin{tabular}{cccc}
        \hline
        \hline
        Instrument & MJD & Magnitude & Band \\
        \hline
        LCO/Sinistro & 59965.32 & $19.35\pm0.01$ & g  \\
        LCO/Sinistro & 59965.32 & $16.65\pm0.01$ & r \\
        LCO/Sinistro & 59965.33 & $15.19\pm0.02$ & i \\
        \hline
        Gaia & 56892.76 & $19.32\pm0.04$ & G \\
        Gaia & 56892.76 & $21.48\pm0.63$ & BP \\
        Gaia & 56892.76 & $17.85\pm0.04$ & RP \\
        Gaia & 56929.55 & $18.94\pm0.01$ & G \\
        Gaia & 56929.55 & $17.41\pm0.03$ & RP \\
        Gaia & 56929.55 & $24.57\pm7.65$ & BP \\
        Gaia & 56929.63 & $18.92\pm0.02$ & G \\
        Gaia & 56929.63 & $17.41\pm0.03$ & RP \\
        Gaia & 56971.00 & $18.02\pm0.01$ & G \\
        Gaia & 56971.00 & $20.14\pm0.22$ & BP \\
        Gaia & 56971.00 & $16.57\pm0.02$ & RP \\
        Gaia & 57025.21 & $16.1\pm0.01$ & G \\
        Gaia & 57054.70 & $15.5\pm0.01$ & G \\
        Gaia & 57054.70 & $17.27\pm0.02$ & BP \\
        Gaia & 57054.70 & $14.2\pm0.01$ & RP \\
        Gaia & 57054.77 & $15.5\pm0.01$ & G \\
        Gaia & 57054.77 & $17.15\pm0.02$ & BP \\
        Gaia & 57054.77 & $14.18\pm0.01$ & RP \\
        Gaia & 57083.02 & $15.01\pm0.01$ & G \\
        Gaia & 57083.02 & $16.79\pm0.01$ & BP \\
        Gaia & 57083.02 & $13.73\pm0.01$ & RP \\
        Gaia & 57126.43 & $16.49\pm0.01$ & BP \\
        Gaia & 57126.43 & $13.41\pm0.01$ & RP \\
        Gaia & 57126.50 & $14.69\pm0.01$ & G \\
        Gaia & 57126.50 & $16.48\pm0.01$ & BP \\
        Gaia & 57126.50 & $13.41\pm0.01$ & RP \\
        Gaia & 57152.43 & $16.54\pm0.01$ & BP \\
        Gaia & 57152.43 & $13.43\pm0.01$ & RP \\
        Gaia & 57152.51 & $16.53\pm0.01$ & BP \\
        Gaia & 57152.51 & $13.43\pm0.01$ & RP \\
        Gaia & 57152.68 & $16.53\pm0.01$ & BP \\
        Gaia & 57152.68 & $13.43\pm0.01$ & RP \\
        Gaia & 57190.92 & $14.66\pm0.01$ & G \\
        Gaia & 57190.92 & $16.44\pm0.01$ & BP \\
        Gaia & 57190.92 & $13.37\pm0.01$ & RP \\
        Gaia & 57225.47 & $14.65\pm0.01$ & G \\
        Gaia & 57225.47 & $16.51\pm0.01$ & BP \\
        Gaia & 57225.47 & $13.36\pm0.01$ & RP \\
        Gaia & 57290.91 & $14.69\pm0.01$ & G \\
        Gaia & 57290.91 & $16.52\pm0.01$ & BP \\
        Gaia & 57290.91 & $13.39\pm0.01$ & RP \\
        Gaia & 57306.09 & $14.69\pm0.01$ & G \\
        Gaia & 57306.09 & $16.53\pm0.01$ & BP \\
        Gaia & 57306.09 & $13.39\pm0.01$ & RP \\
        Gaia & 57306.17 & $14.68\pm0.01$ & G \\
        Gaia & 57306.17 & $16.53\pm0.01$ & BP \\
        Gaia & 57306.17 & $13.38\pm0.01$ & RP \\
        Gaia & 57347.04 & $14.79\pm0.01$ & G \\
        Gaia & 57347.04 & $16.62\pm0.01$ & BP \\
        Gaia & 57347.04 & $13.48\pm0.01$ & RP \\
        Gaia & 57367.04 & $14.85\pm0.01$ & G \\
        Gaia & 57367.11 & $14.86\pm0.01$ & G \\
        Gaia & 57367.11 & $16.75\pm0.01$ & BP \\
        Gaia & 57367.11 & $13.54\pm0.01$ & RP \\
        Gaia & 57401.58 & $14.86\pm0.01$ & G \\
        Gaia & 57401.58 & $16.75\pm0.02$ & BP \\
        Gaia & 57401.58 & $13.56\pm0.01$ & RP \\
        Gaia & 57460.82 & $14.89\pm0.01$ & G \\
        Gaia & 57460.82 & $16.82\pm0.01$ & BP \\
        Gaia & 57460.82 & $13.58\pm0.01$ & RP \\
        \hline
    \end{tabular}
        \caption{Photometric measurements of WTP\,10aaauow.}
    \label{tab:photo_1}
\end{table}

\begin{table}
    \centering
    \contcaption{Photometric measurements of WTP\,10aaauow.}
    \begin{tabular}{cccc}
    \hline
        \hline
        Instrument & MJD & Magnitude & Band \\
        \hline
        Gaia & 57505.48 & $14.97\pm0.01$ & G \\
        Gaia & 57505.48 & $16.87\pm0.01$ & BP \\
        Gaia & 57505.48 & $13.67\pm0.01$ & RP \\
        Gaia & 57505.55 & $14.97\pm0.01$ & G \\
        Gaia & 57505.55 & $16.9\pm0.01$ & BP \\
        Gaia & 57505.55 & $13.66\pm0.01$ & RP \\
        Gaia & 57532.80 & $14.98\pm0.01$ & G \\
        Gaia & 57532.98 & $14.97\pm0.01$ & G \\
        Gaia & 57532.98 & $16.88\pm0.01$ & BP \\
        Gaia & 57532.98 & $13.66\pm0.01$ & RP \\
        Gaia & 57568.21 & $14.96\pm0.01$ & G \\
        Gaia & 57568.21 & $16.73\pm0.01$ & BP \\
        Gaia & 57568.21 & $13.65\pm0.01$ & RP \\
        Gaia & 57568.29 & $14.94\pm0.01$ & G \\
        Gaia & 57568.29 & $13.63\pm0.01$ & RP \\
        Gaia & 57568.29 & $16.85\pm0.01$ & BP \\
        Gaia & 57604.51 & $15.04\pm0.01$ & G \\
        Gaia & 57626.43 & $15.04\pm0.01$ & G \\
        Gaia & 57626.43 & $16.95\pm0.01$ & BP \\
        Gaia & 57626.43 & $13.72\pm0.01$ & RP \\
        Gaia & 57667.95 & $15.06\pm0.01$ & G \\
        Gaia & 57667.95 & $17.02\pm0.01$ & BP \\
        Gaia & 57667.95 & $13.75\pm0.01$ & RP \\
        Gaia & 57668.13 & $15.06\pm0.01$ & G \\
        Gaia & 57668.13 & $17.02\pm0.01$ & BP \\
        Gaia & 57668.13 & $13.74\pm0.01$ & RP \\
        Gaia & 57683.13 & $15.06\pm0.01$ & G \\
        Gaia & 57723.08 & $16.94\pm0.02$ & BP \\
        Gaia & 57723.08 & $13.72\pm0.01$ & RP \\
        Gaia & 57745.15 & $15.09\pm0.01$ & G \\
        Gaia & 57745.15 & $16.98\pm0.02$ & BP \\
        Gaia & 57745.15 & $13.76\pm0.01$ & RP \\
        Gaia & 57778.13 & $15.06\pm0.01$ & G \\
        Gaia & 57778.13 & $17.04\pm0.01$ & BP \\
        Gaia & 57778.13 & $13.75\pm0.01$ & RP \\
        Gaia & 57813.60 & $15.11\pm0.01$ & G \\
        Gaia & 57813.60 & $16.98\pm0.01$ & BP \\
        Gaia & 57813.60 & $13.77\pm0.01$ & RP \\
        Gaia & 57813.67 & $15.1\pm0.01$ & G \\
        Gaia & 57813.67 & $17.01\pm0.01$ & BP \\
        Gaia & 57813.67 & $13.79\pm0.01$ & RP \\
        Gaia & 57839.17 & $15.09\pm0.01$ & G \\
        Gaia & 57839.17 & $17.04\pm0.01$ & BP \\
        Gaia & 57839.17 & $13.78\pm0.01$ & RP \\
        Gaia & 57839.35 & $15.09\pm0.01$ & G \\
        Gaia & 57839.35 & $17.04\pm0.02$ & BP \\
        Gaia & 57839.35 & $13.78\pm0.01$ & RP \\
        Gaia & 57839.42 & $15.1\pm0.01$ & G \\
        Gaia & 57839.42 & $17.05\pm0.01$ & BP \\
        Gaia & 57839.42 & $13.77\pm0.01$ & RP \\
        Gaia & 57883.83 & $15.11\pm0.01$ & G \\
        Gaia & 57883.83 & $17.06\pm0.01$ & BP \\
        Gaia & 57883.83 & $13.78\pm0.01$ & RP \\
        Gaia & 57883.91 & $15.11\pm0.01$ & G \\
        Gaia & 57883.91 & $17.01\pm0.01$ & BP \\
        Gaia & 57883.91 & $13.78\pm0.01$ & RP \\
        \hline
        SOAR/Spartan & 59795.00 & $11.50\pm0.01$ & J \\
        SOAR/Spartan & 59795.00 & $10.20\pm0.01$ & H \\
        SOAR/Spartan & 59795.00 & $9.24\pm0.01$ & Ks \\
        \hline
        2MASS & 51561.87 & $16.19\pm0.11$ & J \\
        2MASS & 51561.87 & $14.77\pm0.07$ & H \\
        2MASS & 51561.87 & $13.81\pm0.05$ & Ks \\
        \hline
    \end{tabular}
    
    \label{tab:photo_2}
\end{table}

\begin{table}
    \centering
    \contcaption{Photometric measurements of WTP\,10aaauow.}
    \begin{tabular}{ccccc}
    \hline
        \hline
        Instrument & MJD & Magnitude & Band \\
        \hline
        Spitzer/IRAC & 53005.53 & $14.39\pm0.01$ & Channel\,1 \\
        Spitzer/IRAC & 53005.53 & $14.18\pm0.01$ & Channel\,2 \\
        Spitzer/IRAC & 53005.53 & $13.69\pm0.01$ & Channel\,3 \\
        Spitzer/IRAC & 53005.53 & $13.64\pm0.01$ & Channel\,4 \\
        \hline
        WISE & 55106.41 & $11.82\pm0.01$ & W1 \\
        WISE & 55106.41 & $10.53\pm0.01$ & W2 \\
        WISE & 55281.51 & $11.66\pm0.01$ & W1 \\
        WISE & 55281.57 & $10.43\pm0.01$ & W2 \\
        WISE & 55349.35 & $8.45\pm0.01$ & W3 \\
        WISE & 55350.72 & $6.15\pm0.01$ & W4 \\
        WISE & 55454.97 & $11.61\pm0.03$ & W1 \\
        WISE & 55454.97 & $10.37\pm0.03$ & W2 \\
        WISE & 56666.86 & $10.27\pm0.03$ & W1 \\
        WISE & 56666.86 & $9.33\pm0.03$ & W2 \\
        WISE & 56844.72 & $10.05\pm0.03$ & W1 \\
        WISE & 56844.72 & $9.16\pm0.03$ & W2 \\
        WISE & 57031.29 & $8.74\pm0.03$ & W1 \\
        WISE & 57031.29 & $7.94\pm0.03$ & W2 \\
        WISE & 57204.97 & $6.98\pm0.12$ & W2 \\
        WISE & 57204.97 & $7.57\pm0.06$ & W1 \\
        WISE & 57395.98 & $7.3\pm0.09$ & W2 \\
        WISE & 57395.98 & $7.94\pm0.03$ & W1 \\
        WISE & 57563.78 & $7.47\pm0.07$ & W2 \\
        WISE & 57563.78 & $8.08\pm0.03$ & W1 \\
        WISE & 57760.18 & $7.57\pm0.06$ & W2 \\
        WISE & 57760.18 & $8.16\pm0.03$ & W1 \\
        WISE & 57924.22 & $7.65\pm0.05$ & W2 \\
        WISE & 57924.22 & $8.24\pm0.03$ & W1 \\
        WISE & 58125.23 & $8.23\pm0.03$ & W1 \\
        WISE & 58125.23 & $7.62\pm0.05$ & W2 \\
        WISE & 58284.66 & $7.68\pm0.05$ & W2 \\
        WISE & 58284.66 & $8.29\pm0.03$ & W1 \\
        WISE & 58491.45 & $7.71\pm0.04$ & W2 \\
        WISE & 58491.45 & $8.33\pm0.03$ & W1 \\
        WISE & 58651.63 & $7.71\pm0.04$ & W2 \\
        WISE & 58651.63 & $8.31\pm0.03$ & W1 \\
        WISE & 58855.65 & $8.27\pm0.03$ & W1 \\
        WISE & 58855.65 & $7.67\pm0.05$ & W2 \\
        WISE & 59015.95 & $7.72\pm0.04$ & W2 \\
        WISE & 59015.95 & $8.29\pm0.03$ & W1 \\
        WISE & 59221.39 & $7.72\pm0.04$ & W2 \\
        WISE & 59221.39 & $8.34\pm0.03$ & W1 \\
        WISE & 59380.14 & $7.78\pm0.04$ & W2 \\
        WISE & 59380.14 & $8.37\pm0.03$ & W1 \\
        WISE & 59586.83 & $7.8\pm0.04$ & W2 \\
        WISE & 59586.83 & $8.45\pm0.03$ & W1 \\
        WISE & 59747.19 & $8.45\pm0.03$ & W1 \\
        WISE & 59747.19 & $7.82\pm0.04$ & W2 \\
        \hline
    \end{tabular}
    \label{tab:photo_3}
\end{table}


\bsp	
\label{lastpage}
\end{document}